\documentclass[
	a4paper, 
	10pt, 
	unnumberedsections, 
	twoside, 
	twocolumn,
]{article}
\usepackage{etoolbox} 

\newtoggle{unnumberedsections} 
\settoggle{unnumberedsections}{false} 
\relax 

\usepackage{graphicx} 
\graphicspath{{Figures/}{./}} 

\usepackage[bottom, hang]{footmisc} 
\usepackage[
	top=2.5cm, 
	bottom=2.5cm, 
	left=2cm, 
	right=2cm, 
	footskip=1cm, 
	headsep=0.75cm, 
	columnsep=20pt, 
]{geometry}

\usepackage[utf8]{inputenc} 
\usepackage[T1]{fontenc} 

\usepackage[sc]{mathpazo} 

\usepackage{microtype} 

\usepackage{fancyhdr} 
\usepackage{titlesec} 

\iftoggle{unnumberedsections}{ 
	\setcounter{secnumdepth}{0} 
}{
	\setcounter{secnumdepth}{2} 
}

\titleformat
	{\section} 
	[block] 
	{\Large\bfseries\centering} 
	{\thesection} 
	{0.5em} 
	{} 
	[] 

\titleformat
	{\subsection} 
	[block] 
	{\raggedright\large\bfseries} 
	{\thesubsection} 
	{0.5em} 
	{} 
	[] 

\titleformat
	{\subsubsection} 
	[runin] 
	{\bfseries} 
	{} 
	{5pt} 
	{} 
	[] 

\titlespacing*{\subsubsection}{0pt}{0.5\baselineskip}{8pt} 

\usepackage{titling} 

\pretitle{\begin{center}\huge\bfseries} 
\posttitle{\end{center}} 

\patchcmd{\maketitle}{plain}{empty}{}{} 

\usepackage{abstract} 

\usepackage[
	backend=biber, 
	citestyle=numeric, 
	bibstyle=numeric, 
	sorting=none, 
]{biblatex}

\usepackage{booktabs} 

\usepackage{array} 

\newcolumntype{R}[1]{>{\raggedleft\arraybackslash}p{#1}} 
\newcolumntype{L}[1]{>{\raggedright\arraybackslash}p{#1}} 
\newcolumntype{C}[1]{>{\centering\arraybackslash}p{#1}} 

\usepackage{caption} 

\usepackage{enumitem} 

\setlist{noitemsep} 

\usepackage{hyperref} 

\hypersetup{
	colorlinks=true, 
	urlcolor=black, 
	linkcolor=black, 
	citecolor=black, 
}

\newcommand{\runninghead}[1]{\renewcommand{\runninghead}{#1}}

\newcommand{\footertext}[1]{\renewcommand{\footertext}{#1}}

\usepackage{amsmath}
\usepackage{amssymb}
\usepackage{mathtools}
\usepackage{placeins}

\newcommand{\blue}[1]{\color{black}{#1}\color{black}}
\newcommand{\green}[1]{\color{black}{#1}\color{black}}
\newcommand{\red}[1]{\color{black}{#1}\color{black}}

\title{A Comparison of Rosenbrock-Wanner and Crank-Nicolson Time Integrators for Atmospheric Modelling}
\author{%
	David Lee\textsuperscript{1}\thanks{Corresponding author: \href{mailto:david.lee@bom.gov.au}{david.lee@bom.gov.au}}
}

\date{\footnotesize\textsuperscript{\textbf{1}}Bureau of Meteorology, Melbourne, Australia}

\renewcommand{\maketitlehookd}{%
\begin{abstract}
Non-hydrostatic atmospheric models often use semi-implicit temporal discretisations in order
to negate the time step limitation of explicitly resolving the fast acoustic and gravity 
waves. Solving the resulting system to machine precision using Newton's method is considered 
prohibitively expensive, and so the non-linear solver is typically truncated to a fixed 
number of iterations, using an approximate Jacobian matrix that is reassembled only once 
per time step. The present article studies the impact of using various third-order, 
four stage Rosenbrock-Wanner schemes, where integration weights are chosen to meet specific 
stability and order conditions, in comparison to a Crank-Nicolson time discretisation, 
as is done in the UK Met Office's \emph{LFRic} model. Rosenbrock-Wanner schemes 
present a promising alternative on account of their ability to preserve their temporal 
order with only an approximate Jacobian, and may be constructed to be stiffly-stable, 
so as to ensure the decay of fast unresolved modes. These schemes are compared
for the 2D rotating shallow water equations and the 3D compressible Euler equations at 
both planetary and non-hydrostatic scales and are shown to exhibit improved results in terms
of their energetic profiles and stability.
Results in terms of computational performance are mixed, with the Crank-Nicolson method
allowing for longer time steps and faster time to solution for the baroclinic instability test 
case at planetary scales, and the Rosenbrock-Wanner methods allowing for longer time steps and 
faster time to solution for a rising bubble test case at non-hydrostatic scales.
\end{abstract}
}

\begin{document}

\maketitle

Semi-implicit time discretisations are a popular choice for non-hydrostatic atmospheric models, 
since they negate the time step limitation associated with the explicit solution of acoustic 
and gravity waves \cite{Melvin et. al. (2019), Maynard et. al. (2020)}. This is particularly true for finite element spatial discretisations using
Gauss-Lobatto quadrature rules, since unlike spectral element or discontinuous Galerkin methods 
using inexact spatial integration, these finite element methods have non-diagonal mass matrices
which require the implicit solution of linear systems even for the case of explicit time 
integration schemes. These implicit non-linear systems are commonly solved using Newton's method,
for which the residual vector is discretised using a second order, time centered (or off-centered)
Crank-Nicolson scheme, or some similar iterative method. However due to computational performance
limitations, the non-linear solver is typically truncated to a fixed number of iterations, rather
than to the convergence of the residual below some specified tolerance. Moreover in order to further
improve computational performance, the Jacobian operator used to determine the descent direction 
at each iteration is often approximated by omitting terms 
\green{that are not strongly related to the fast wave dynamics}, 
the use of vertical reference profiles, and the re-use of the same Jacobian between 
non-linear iterations or even time steps. Consequently this approximate solution may be regarded as a 
quasi-Newton method, rather than a full Newton method for which the Jacobian would incorporate all 
derivatives of the residual vector with respect to all solution variables, and be re-assembled at 
each non-linear iteration.

As an alternative to such a scheme, here we consider the use of Rosenbrock-Wanner (RoW) methods, 
where a finite number of non-linear iterations are replaced by a fixed number of implicit Runge-Kutta
stages, for which the weights are chosen so as to satisfy specific order and stability conditions \cite{Hairer and Wanner (1996), Rang and Angermann (2005)}.
In the current study we limit ourselves to four stage schemes 
\red{since this is the same number of iterations used as a default in the Crank-Nicolson (CN) quasi-Newton 
scheme in the UK Met Office's \emph{LFRic} dynamical core \cite{Melvin et. al. (2019)}, the model we use to compare these 
schemes in the present study. }
The four-stage, third order RoW schemes considered here are for the most part stiffly-stable methods which 
are well suited to geophysical systems that involve fast acoustic and gravity waves.

\green{The 3D compressible Euler equations used to simulate the dynamics of a dry atmosphere include both 
prognostic equations for the velocity, the density and the thermodynamic variable, and an algebraic equation, 
in the form of } an equation of state. 
\green{Block factorisation of the full coupled system, together with a lumped approximation to the velocity space mass matrix, }
leads to a non-singular Helmholtz equation for the pressure \cite{Maynard et. al. (2020)}.
\green{This resultant Helmholtz equation may be used as an
approximate preconditioner for the original coupled system of equations. }
\red{In order to recover the original coupled system from this Helmholtz equation (ignoring complications with the spatial
discretisation due to mass lumping), one would have to differentiate twice in time. Such equations are known as "index-2"
partial differential algebraic equations (PDAEs). }
Consequently the RoW methods studied here must also support index-2 PDAEs \green{in order to allow for the
use of the Helmholtz pressure equation as an approximate preconditioner for the full coupled system. }

RoW methods (for which only an approximate Jacobian is required) have shown 
promise for the solution of the incompressible Navier-Stokes equations in two \cite{John et. al. (2006)} and 
three \cite{Deparis et. al. (2019)} spatial dimensions, which also involve an index-2 PDAE elliptic equation, 
in the form of a Poisson equation for the pressure. Therefore it is reasonable to expect that they would 
also be beneficial for the solution of geophysical systems. Another advantage of RoW schemes 
is that they posses an embedded low order solution, which may be used for adaptive time step control 
\cite{John and Rang (2010)}. They have also been applied to the simulation of compressible Navier-Stokes in two
\cite{Blom et. al. (2016)} and three \cite{Liu et. al. (2016)} dimensions, as well as to optimal control problems \cite{Lang and Verwer (2013)}.
A second order, three stage Rosenbrock scheme has also been used in the ASAM atmospheric model \cite{Jahn et. al. (2015)}.

\green{RoW schemes are closely related to diagonally implicit Runge-Kutta (DIRK) methods 
\cite{Kennedy and Carpenter (2019)}, and have shown improved performance with respect to DIRK schemes for compressible
flows where the RoW method Jacobian is only assembled once per time step \cite{Liu et. al. (2016)}. }
\blue{While RoW schemes are somewhat flexible as to the approximation of the 
Jacobian, they are prone to order reduction for stiff systems \cite{Rang and Angermann (2005), Rang (2015)}. }
\red{Since the simulation results will be somewhat sensitive to the choice of Jacobian approximation 
for both the RoW and CN method, we present results for both the rotating shallow water 
and 3D compressible Euler equations, using compatible finite element methods with different polynomial
degrees and representations of the transport terms. }

\blue{The purpose of the present study is to compare the RoW and CN schemes, firstly in terms 
of their energetic profiles, since this gives some insight into their
ability to faithfully represent dynamical processes and also the presence of spurious computational 
artefacts that may cause stability issues when coupling to external forcing or source terms with similar 
time scales. Secondly we compare the schemes in terms of time to solution for their maximum stable
time step sizes in order to benchmark their computational efficiency. We also present some analysis 
detailing the similarities of the RoW and CN schemes, and how variants of the CN scheme may be constructed
to enhance its stability properties for fast modes. A new sub-stepping temporal scheme is also presented 
for the transport terms for the RoW scheme, so that these can be applied in a more computationally efficient 
manner. }

The remainder of this article proceeds as follows: In Section 1 the formulation of RoW 
methods will be briefly introduced. More detailed discussions can be found in the references therein.
Section 2 describes the geophysical systems studied in this article, namely the 2D rotating shallow
water equations and the 3D compressible Euler equations. Results comparing the application of 
four stage RoW methods to a CN scheme for standard test 
cases for these systems will be presented in Section 3. Finally, conclusions based on these results
will be presented in Section 4.

\section{Introduction to Rosenbrock-Wanner methods}

We are concerned with systems of PDAEs with a temporal structure of the form
\begin{subequations}
\begin{align}
	\boldsymbol{\mathsf{M}}\frac{d\boldsymbol{x}}{dt} &= F(\boldsymbol{x},\boldsymbol{y}) \\
	\boldsymbol{0} &= G(\boldsymbol{x},\boldsymbol{y})
\end{align}
\end{subequations}
where $\boldsymbol{\mathsf{M}}$ is some mass matrix \green{(which may be strictly diagonal for a finite difference
or a finite volume scheme)}, $\boldsymbol{x}$, $\boldsymbol{y}$ are vectors
of prognostic and diagnostic state variables respectively, $F$ includes all forcing terms for the 
prognostic equations and $G$ is a set of time independent algebraic equations\red{, and $d/dt$ is an
Eulerian derivative in time only}. If $F$ and $G$ are 
non-linear functions of $\boldsymbol{x}$ and $\boldsymbol{y}$ then we may apply a centered CN
temporal discretisation with respect to a time step $\Delta t$ and solve the resulting system for 
time step $n+1$ using Newton's method at each non-linear iteration $i\ge 1$ with an initial state for the
time step $\boldsymbol{y}^0=\boldsymbol{y}^n$ as
\begin{multline}\label{eq::cn}
\begin{bmatrix}
	\boldsymbol{\mathsf{M}} - \gamma\Delta t \boldsymbol{\mathsf{W}}_{F,x}^{(i)} &
	- \gamma\Delta t \boldsymbol{\mathsf{W}}_{F,y}^{(i)} \\
	- \boldsymbol{\mathsf{W}}_{G,x}^{(i)} &
	- \boldsymbol{\mathsf{W}}_{G,y}^{(i)} 
\end{bmatrix}
	\begin{bmatrix}
		\delta\boldsymbol{x}^{(i)} \\
		\delta\boldsymbol{y}^{(i)} 
	\end{bmatrix} = \\
	\begin{bmatrix}
	\boldsymbol{\mathsf{M}}(\boldsymbol{x}^n - \boldsymbol{x}^{(i-1)}) + 
	\frac{\Delta t}{2}\Big(F(\boldsymbol{x}^n,\boldsymbol{y}^n) + F(\boldsymbol{x}^{(i-1)},\boldsymbol{y}^{(i-1)})\Big) \\
	G(\boldsymbol{x}^{(i-1)},\boldsymbol{y}^{(i-1)})
	\end{bmatrix} 
\end{multline}
where $\gamma = 1/2$ for an optimal rate of convergence (\green{or $\gamma > 1/2$ for an over-relaxed solution}), and
\begin{equation}
	\boldsymbol{\mathsf{W}}^{(i)}=
	\begin{bmatrix}
		\boldsymbol{\mathsf{W}}_{F,x}^{(i)} &
		\boldsymbol{\mathsf{W}}_{F,y}^{(i)} \\
		\boldsymbol{\mathsf{W}}_{G,x}^{(i)} &
		\boldsymbol{\mathsf{W}}_{G,y}^{(i)} 
	\end{bmatrix}
	\approx
	\begin{bmatrix}
		\frac{\delta F(\boldsymbol{x}^{(i)},\boldsymbol{y}^{(i)})}{\delta\boldsymbol{x}^{(i)}} &
		\frac{\delta F(\boldsymbol{x}^{(i)},\boldsymbol{y}^{(i)})}{\delta\boldsymbol{y}^{(i)}} \\
		\frac{\delta G(\boldsymbol{x}^{(i)},\boldsymbol{y}^{(i)})}{\delta\boldsymbol{x}^{(i)}} &
		\frac{\delta G(\boldsymbol{x}^{(i)},\boldsymbol{y}^{(i)})}{\delta\boldsymbol{y}^{(i)}} 
	\end{bmatrix}
\end{equation}
is the (approximate) Jacobian matrix evaluated at iteration $i$ 
\red{(with parenthesis, $(\cdot)$ being used to denote the iteration index as opposed to the time step) } and
\begin{subequations}\label{eq::newton_recon}
	\begin{align}
		\boldsymbol{x}^{(i+1)} &= \boldsymbol{x}^{(i)} + \delta\boldsymbol{x}^{(i)} =
		\boldsymbol{x}^n + \sum_{j=1}^i\delta\boldsymbol{x}^{(j)},\\
		\boldsymbol{y}^{(i+1)} &= \boldsymbol{y}^{(i)} + \delta\boldsymbol{y}^{(i)} =
		\boldsymbol{y}^n + \sum_{j=1}^i\delta\boldsymbol{y}^{(j)}.
	\end{align}
\end{subequations}
The iteration is terminated and the solution at time level $n+1$ is updated once 
$||\delta\boldsymbol{x}^{(i)}||$, $||\delta\boldsymbol{y}^{(i)}||$ are below some specified tolerance.

As discussed above, solving to \red{machine precision } and re-assembling $\boldsymbol{\mathsf{W}}$ at each iteration 
$i$ is prohibitively expensive for many applications, and so typically this is assembled only once per 
time step as $\boldsymbol{\mathsf{W}}^{(1)}$. Also rather than terminate when the solution increments are 
below some tolerance, the stopping condition is set for a fixed number of iterations $s$ as $i > s$.

In contrast to the Newton iteration with the CN time discretisation described above, the
RoW method with $s$-stages is given for an autonomous system of PDAEs (where $F$ and $G$ do 
not depend directly on $t$) at iteration $1\le i\le s$ as \cite{Hairer and Wanner (1996), Rang and Angermann (2005), Rang (2013)}
\begin{multline}\label{eq::ros}
\begin{bmatrix}
	\boldsymbol{\mathsf{M}} - \gamma\Delta t \boldsymbol{\mathsf{W}}_{F,x}^{(1)} &
	- \gamma\Delta t \boldsymbol{\mathsf{W}}_{F,y}^{(1)} \\
	- \boldsymbol{\mathsf{W}}_{G,x}^{(1)} &
	- \boldsymbol{\mathsf{W}}_{G,y}^{(1)} 
\end{bmatrix}
	\begin{bmatrix}
		\boldsymbol{k}^{(i)} \\
		\boldsymbol{l}^{(i)} 
	\end{bmatrix} = \\
	\begin{bmatrix}
		\Delta tF(\boldsymbol{x}_*^{(i)},\boldsymbol{y}_*^{(i)}) \\
		G(\boldsymbol{x}_*^{(i)},\boldsymbol{y}_*^{(i)}) 
	\end{bmatrix} + \\
\begin{bmatrix}
	\gamma\Delta t \boldsymbol{\mathsf{W}}_{F,x}^{(1)} &
	\gamma\Delta t \boldsymbol{\mathsf{W}}_{F,y}^{(1)} \\
	\boldsymbol{\mathsf{W}}_{G,x}^{(1)} &
	\boldsymbol{\mathsf{W}}_{G,y}^{(1)} 
\end{bmatrix}
	\sum_{j=1}^{i-1}\gamma_{ij}
	\begin{bmatrix}
		\boldsymbol{k}^{(j)} \\
		\boldsymbol{l}^{(j)} 
	\end{bmatrix} 
\end{multline}
where
\begin{equation}\label{eq::ros_x_star}
	\boldsymbol{x}_*^{(i)} = \boldsymbol{x}^n + \sum_{j=1}^{i-1}\alpha_{ij}\boldsymbol{k}^{(j)},\qquad
	\boldsymbol{y}_*^{(i)} = \boldsymbol{y}^n + \sum_{j=1}^{i-1}\alpha_{ij}\boldsymbol{l}^{(j)}
\end{equation}
for the scalar weights $\alpha_{ij}$ and $\gamma_{ij}$. Unlike Newton's method, where
the solution at the end of the time step is just the previous solution plus a sum over all the
solution increments $\delta\boldsymbol{x}^{(j)}$, $\delta\boldsymbol{y}^{(j)}$ \eqref{eq::newton_recon}, 
the solution at time level $n+1$ for the RoW method is reconstructed from the 
scalar weights $b_i$ 
\blue{and the solutions $\boldsymbol{k}^{(i)}$, $\boldsymbol{l}^{(i)}$ for the prognostic and diagnostic increments respectively at stage $i$ } as
\begin{equation}
	\boldsymbol{x}^{n+1} = \boldsymbol{x}^n + \sum_{i=1}^sb_i\boldsymbol{k}^{(i)},\qquad
	\boldsymbol{y}^{n+1} = \boldsymbol{y}^n + \sum_{i=1}^sb_i\boldsymbol{l}^{(i)}.
\end{equation}
In order to ensure that the left hand side operator only needs to be assembled once per time
step, the weights are customarily chosen such that $\gamma = \gamma_{ii}$ is constant for all 
stages $i$.

\green{Any external forcing terms due to physics parameterisations or coupling to other model 
components may be added to the RoW scheme at stage $i$ and time $t^n + \sum_{j}\alpha_{ij}\Delta t$
via the augmentation of $\boldsymbol{x}_*^{(i)}$ in \eqref{eq::ros_x_star}.}

RoW methods are a sub-class of Rosenbrock methods for which the coefficients 
$\alpha_{i,j}$, $\gamma_{i,j}$ are chosen to satisfy the required order condition for only
an approximate representation of the Jacobian, $\boldsymbol{\mathsf{W}}$. For some classes
of RoW methods, such as Krylov-ROW \cite{Schmitt and Weiner (1995), Weiner et. al. (1997)} and Rosenbrock-Krylov 
methods \cite{Tranquilli and Sandu (2014)}, this approximate Jacobian is derived from a low-rank approximation to 
the Krylov subspace generated from the actual Jacobian (which is often constructed via a 
matrix-free differencing of the residual vector). In the present context, this approximate
Jacobian is constructed \cite{Melvin et. al. (2019), Maynard et. al. (2020)} 
\green{in order to treat the terms leading to fast wave dynamics in a semi-implicit fashion, } via the omission of non-stiff terms
and the \blue{use of constant thermodynamic profiles over all nonlinear iterations within a single time step}.

\subsection{Quasi-Newton Crank-Nicolson time discretisation as a Rosenbrock-Wanner scheme}\label{sec::1.1}

In order to further illustrate the comparison between a finite iteration quasi-Newton method and 
RoW methods, we show that the two stage, second order ROS2 scheme \cite{Verwer et. al. (1999)} 
is equivalent to two iterations of a Newton method with a CN time discretisation. For 
the sake of brevity, we show this equivalence for a system of equations involving prognostic 
equations only, however this equivalence also holds with the inclusion of algebraic equations.

Two iterations of the CN scheme \eqref{eq::cn} give a solution as
\begin{subequations}\label{eq::cn_2stage}
	\begin{align}
		(\boldsymbol{\mathsf{M}} - \gamma\Delta t\boldsymbol{\mathsf{W}}^{(1)})\delta\boldsymbol{x}^{(1)} &= 
		\Delta tF(\boldsymbol{x}^n),\label{eq::cn_ros2_it1}\\
		\boldsymbol{x}^{(1)} &= \boldsymbol{x}^n +\delta\boldsymbol{x}^{(1)},\\
		(\boldsymbol{\mathsf{M}} - \gamma\Delta t\boldsymbol{\mathsf{W}}^{(1)})\delta\boldsymbol{x}^{(2)} &= 
		-\boldsymbol{\mathsf{M}}\delta\boldsymbol{x}^{(1)} + \notag \\ 
		\frac{\Delta t}{2}(F(\boldsymbol{x}^n) + F(\boldsymbol{x}^{(1)})),\label{eq::cn_ros2_it2}\\
		\boldsymbol{x}^{n+1} &= \boldsymbol{x}^n +\delta\boldsymbol{x}^{(1)} + \delta\boldsymbol{x}^{(2)}
	\end{align}
\end{subequations}
In order to show the equivalence of the second iteration of the two iteration CN scheme 
\eqref{eq::cn_ros2_it2} to the second stage of the ROS2 scheme (where $\alpha_{21}=1$, $\gamma_{21}=-2$, 
$b_1=b_2=1/2$) we 
apply the substitution
$\boldsymbol{k}^{(1)} = \delta\boldsymbol{x}^{(1)}$, $\boldsymbol{k}^{(2)} = 2\delta\boldsymbol{x}^{(2)} + \delta\boldsymbol{x}^{(1)}$ 
and construct the solution using the integration weights
$\boldsymbol{x}^{n+1} = \boldsymbol{x}^n + \boldsymbol{k}^{(1)}/2 + \boldsymbol{k}^{(2)}/2$.
While the above scheme is 
typically run with $\gamma=1/2$, which is neutrally stable \red{(for linear problems)}, an L-stable variant is given as 
$\gamma=1\pm\sqrt{2}/2$ \cite{Verwer et. al. (1999)}.
\red{L-stability ensures that for linear problems the amplitudes of the largest eigenvalues decay to zero. We consider
this to be a desirable property for compressible atmospheric solvers since this should help to mitigate against unresolved 
fast temporal oscillations, as will be discussed in Section \ref{sec::bw}.}

We can extend this comparison to a quasi-Newton method with any number of stages, for which we have
at iteration $i>1$ that $\delta\boldsymbol{x}^{(i)} = \boldsymbol{k}^{(i)}/2 - \boldsymbol{k}^{(i-1)}/2$. For 
the four iteration Crank-Nicolson scheme (CN4), this is given as an equivalent RoW 
scheme for the matrices $\mathsf{\Gamma}$, $\mathsf{A}$ as
\begin{subequations}
	\begin{align}
	\mathsf{\Gamma} &= (\gamma_{i,j})_{i,j=1}^4 =
	\gamma\begin{bmatrix}
		1 & 0 & 0 & 0 \\
		-2 & 1 & 0 & 0 \\
		-1 & -1 & 1 & 0 \\
		-1 & 0 & -1 & 1 
	\end{bmatrix},\\
	\mathsf{A} &= \begin{bmatrix}(\alpha_{i,j})_{i,j=1}^{3,4}\end{bmatrix} = 
	\begin{bmatrix}
		0 & 0 & 0 & 0 \\
		1 & 0 & 0 & 0 \\
		\frac{1}{2} & \frac{1}{2} & 0 & 0 \\
		\frac{1}{2} & 0 & \frac{1}{2} & 0
	\end{bmatrix},\\
	\boldsymbol{b} &= \begin{bmatrix}\frac{1}{2} & 0 & 0 & \frac{1}{2}\end{bmatrix}.
	\end{align}
\end{subequations}

For the prototypical case where $d\boldsymbol{x}/dt = \lambda\boldsymbol{x}$, 
$\boldsymbol{\mathsf{W}} = \delta F(\boldsymbol{x})/\delta\boldsymbol{x}=\lambda$
we have that $\boldsymbol{x}^{n+1} = R(\Delta t\lambda)\boldsymbol{x}^n$ where
$R(\Delta t\lambda) = 1 + \Delta t\lambda\boldsymbol{b}^{\top}(\mathsf{I} - \Delta t\lambda\mathsf{B})^{-1}\boldsymbol{1}$ 
is the amplification factor (see \cite{Hairer and Wanner (1996)}, ch. IV.7), $\mathsf{I}$ is the identity matrix and 
$\mathsf{B} = \mathsf{\Gamma} + \mathsf{A}$. For $\gamma = 1/2$ we have that 
$R(\Delta t\lambda) = (1+\Delta t\lambda/2)/(1-\Delta t\lambda/2)$, such that the solution is 
neutrally stable \red{in the linear regime } with eigenvalues on the unit circle, as is the case for the two stage 
CN2/ROS2 scheme \eqref{eq::cn_2stage} \cite{Verwer et. al. (1999)}.
More generally we have the amplification factor for the four stage CN Newton method, using
$z=\Delta t\lambda$ for brevity, as

\begin{multline}
R(z) = \frac{-z}{\gamma z-1} + \frac{z^2(0.5-\gamma)}{2(\gamma z-1)^2} + \frac{z^3(-\gamma^2 +\gamma -0.25)}{2(\gamma z-1)^3} + \\
	\Bigg(z^2(-4\gamma^3z^2+4.5\gamma ^2z^2+3\gamma^2z-1.75gz^2-2\gamma z-\gamma + \\ 0.25z^2+0.25z+0.5)\Bigg)\Bigg/2(\gamma z-1)^4 + 1.
\end{multline}
The coefficient of the $z^4$ term in the numerator, 
$\gamma^4 - 4\gamma^3 + 3\gamma^2 - \gamma + 0.125$,
has real roots as $\gamma=0.2716068084314726$ and $\gamma=3.1426067539416227$.
\green{For these values the numerator will then be one polynomial degree lower than the denominator, 
such that by l'H\^opital's rule the amplification factor will go to zero as $z\rightarrow\infty$, leading 
to L-stability. These L-stable variants of the CN scheme were tested for the 3D compressible
Euler equations in the baroclinic instability configuration described in Section \ref{sec::bw_time}. However
only the $\gamma=3.1426$ variant was observed to be stable, and only for the case of upwinded transport  
and for time steps much shorter than the standard scheme.}

\begin{figure}[!hbtp]
\begin{center}
\includegraphics[width=0.48\textwidth,height=0.36\textwidth]{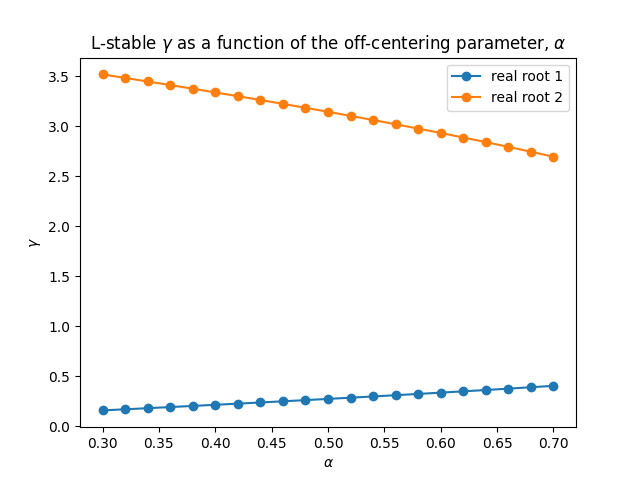}
\caption{L-stable $\gamma$ values for the off-centered CN scheme.}
\label{fig::L-stable_gamma}
\end{center}
\end{figure}

For the case of an off-centered CN time integration scheme, for which the right hand side
in \eqref{eq::cn} is replaced with
\begin{multline}\label{eq::cn_rhs_alpha}
	\boldsymbol{\mathsf{M}}(\boldsymbol{x}^n - \boldsymbol{x}^{(i-1)}) + \\
	\Delta t\Bigg((1-\alpha)F(\boldsymbol{x}^n,\boldsymbol{y}^n) + \alpha F(\boldsymbol{x}^{(i-1)},\boldsymbol{y}^{(i-1)})\Bigg)
\end{multline}
for constant $\alpha$, we have a recurrence relation of 
$\delta\boldsymbol{x}^{(i)} = \alpha (\boldsymbol{k}^{(i)} - \boldsymbol{k}^{(i-1)})$,
$\boldsymbol{x}^{(i)} = \alpha\boldsymbol{k}^{(i)} + (1-\alpha)\boldsymbol{k}^{(1)}$. The corresponding 
matrix expressions are given as

\begin{subequations}
	\begin{align}
		\mathsf{\Gamma} &= (\gamma_{i,j})_{i,j=1}^4 =
	\gamma\begin{bmatrix}
		1 & 0 & 0 & 0 \\
		\frac{-1}{\alpha} & 1 & 0 & 0 \\
		\frac{\alpha-1}{\alpha} & -1 & 1 & 0 \\
		\frac{\alpha-1}{\alpha} & 0 & -1 & 1 
	\end{bmatrix},\\
		\mathsf{A} &= \begin{bmatrix}(\alpha_{i,j})_{i,j=1}^{3,4}\end{bmatrix} = 
	\begin{bmatrix}
		0 & 0 & 0 & 0 \\
		1 & 0 & 0 & 0 \\
		1-\alpha & \alpha & 0 & 0 \\
		1-\alpha & 0 & \alpha & 0
	\end{bmatrix},\\
		\boldsymbol{b} &= \begin{bmatrix}1-\alpha & 0 & 0 & \alpha\end{bmatrix}.
	\end{align}
\end{subequations}
For this general case the coefficient of the $z^4$ term are
$\gamma^4 - 4\gamma^3 + 6\alpha\gamma^2 - 4\alpha^2\gamma + \alpha^3$. As observed in Fig. 
\ref{fig::L-stable_gamma}, the real roots of this expression vary only weakly \green{with } $\alpha$.
\green{Via the binomial series, the leading error for the centered CN scheme is 
$|\frac{1}{12}z^3|$. For the off-centered case \eqref{eq::cn_rhs_alpha} this increases to 
$|(\frac{1}{2}-\alpha)z^2|$, and the method is only first order accurate for
$|\frac{1}{2}-\alpha|>|{z}^{-1}|$, such that the accuracy degrades most rapidly large 
eigenvalues and time steps.}

\section{Geophysical systems}

This article considers two different geophysical systems, the two dimensional rotating shallow
water equations on the sphere, and the three dimensional compressible Euler equations on both 
spherical and planar geometry. The rotating shallow water equations are a widely used prototypical
model of geophysical phenomena on account of their capacity to represent 
\green{both fast and slow nonlinear dynamics due to gravity and Rossby waves respectively. }
In the present case, comparing CN and RoW schemes
for the rotating shallow water equations allows us to benchmark the allowable time step and 
conservation properties of these methods for the case of a very simple approximate Jacobian at 
low mach-number.
\red{The results for both the RoW and CN
simulations will be somewhat sensitive to the choices made in the approximation of the Jacobian, so applying these
methods to model configurations with different spatial discretisations and dynamics will help to determine commonalities
between the results.}

\subsection{The rotating shallow water equations}

The two dimensional rotating shallow water equations are given for the velocity $\boldsymbol{u}$ 
and the fluid depth $h$ as
\begin{subequations}\label{eq::rsw}
	\begin{alignat}{3}
	\frac{\partial\boldsymbol{u}}{\partial t} &= 
	-q\times\boldsymbol{U} - \nabla\Phi
		&&\blue{:= } F_u(\boldsymbol{u},h)\\
	\frac{\partial h}{\partial t} &= 
		-\nabla\cdot\boldsymbol{U} &&\blue{:= } F_h(\boldsymbol{u},h)
\end{alignat}
\end{subequations}
where $f$ is the Coriolis term due to the earth's rotation, $g$ is (constant) gravity, 
\red{$\boldsymbol{U}:= h\boldsymbol{u}$ is the mass flux, $\Phi := \frac{1}{2}\boldsymbol{u}\cdot\boldsymbol{u} + gh$
is the Bernoulli potential and $q:=(\nabla\times\boldsymbol{u} + f)/h$ is the potential vorticity. } For a mean fluid
depth $H$ much greater than the variations in $h$ and for spatial and temporal scales consistent
with the earth's rotation, the above system may be solved 
using a constant in time approximate
Jacobian of the form \cite{Bauer et. al. (2018), Wimmer et. al. (2020), Lee et. al. (2022)}
\begin{equation}
	\boldsymbol{\mathsf{W}}_{rsw} = 
	\begin{bmatrix}\boldsymbol{\mathsf{C}} & -g\boldsymbol{\mathsf{G}} \\
		H\boldsymbol{\mathsf{G}}^{\top} & \boldsymbol{\mathsf{0}}
	\end{bmatrix}
	\approx
	\begin{bmatrix}
		\frac{\partial F_u^h}{\partial\boldsymbol{u}^h} & 
		\frac{\partial F_u^h}{\partial h^h} \\ 
		\frac{\partial F_h^h}{\partial\boldsymbol{u}^h} & 
		\frac{\partial F_h^h}{\partial h^h} 
	\end{bmatrix},
\end{equation}
where $\boldsymbol{\mathsf{C}}$ is the Coriolis operator and $\boldsymbol{\mathsf{G}}$ is the gradient 
operator (for which the divergence operator is assumed to be its adjoint assuming periodic boundary conditions). 
The approximate Jacobian above omits non-linear terms associated with both mass and momentum transport, and
instead assumes a linearisation around a state of constant mean fluid depth, planetary rotation and 
gravitational potential. \green{Note that for the 3D compressible Euler equations described in the proceeding
section, the presence of a complex and varying thermodynamic profile and acoustic modes means that the 
approximate Jacobian must be periodically reassembled, and incorporate additional terms not required for the 
shallow water case at low Mach number. }

The shallow water equations consist of only prognostic equations for which 
$\boldsymbol{x}_h=(\boldsymbol{u}_h,h_h)$ is the full state vector and there is no $\boldsymbol{y}_h$ vector of
algebraic constraints. 
\red{Note the use of subscripts, $h$, to denote discrete representations of solution 
variables and forcing terms defined within an appropriate finite element subspace. } 
\red{In order to derive the discrete variational form of the equations from the continuous system 
\eqref{eq::rsw} via the Galerkin method, one approximates
the solution variables $\boldsymbol{u}$, $h$ by finite element trial functions which span the discrete
equivalent of the $H(div)$ and $L^2$ spaces respectively, and then multiples the momentum and continuity
equations by test functions of the same form (applying integration by parts to derive a weak gradient of $\Phi_h$). 
The discrete forms of $\boldsymbol{U}_h$, $\Phi_h$ and $q_h$ 
are also determined via Galerkin projections at each nonlinear iteration into the discrete subspaces of 
$H(div)$, $L^2$ and $H^1$ respectively. } 
See the above references for the specific definition of these operators using a mixed
finite element discretisation of the rotating shallow water equations. 

\subsection{The 3D compressible Euler equations}

While the rotating shallow water equations are a 
\green{useful system for studying the suitability of numerical methods for representing gravity waves and
barotropic turbulence, }
to fully capture the dynamics of a dry atmosphere at both planetary and non-hydrostatic scales, we also study
RoW time integration for the 3D compressible Euler equations under the shallow 
atmosphere approximation, given for the velocity $\boldsymbol{u}$, density $\rho$, potential temperature
$\theta$ and Exner pressure $\Pi$ as \cite{Melvin et. al. (2019), Maynard et. al. (2020), Lee (2021), Lee and Palha(2021)} as
\begin{subequations}\label{eq::ce}
	\begin{alignat}{3}
	\frac{\partial\boldsymbol{u}}{\partial t} &= 
		-(\nabla\times\boldsymbol{u} + f)\times\boldsymbol{u}\notag\\ &- \nabla\Bigg(\frac{\boldsymbol{u}\cdot\boldsymbol{u}}{2} + gz\Bigg)
		-c_p\theta\nabla\Pi &&\blue{:= }
	F_u(\boldsymbol{u},\rho,\theta,\Pi)\label{eq::mom}\\
	\frac{\partial\rho}{\partial t} &= 
		-\nabla\cdot(\rho\boldsymbol{u}) &&\blue{:= } F_{\rho}(\boldsymbol{u},\rho)\label{eq::mass}\\
	\frac{\partial\theta}{\partial t} &= 
		-\boldsymbol{u}\cdot\nabla\theta &&\blue{:= } F_{\theta}(\boldsymbol{u},\theta)\label{eq::temp}\\
		0 &= \Pi - \Bigg(\frac{R\rho\theta}{p_0}\Bigg)^{\frac{R}{c_v}} &&\blue{:= } G_{\Pi}(\rho,\theta,\Pi),\label{eq::eos}
\end{alignat}
\end{subequations}
where $z$ is the vertical coordinate, $c_p$ and $c_v$ are the specific heats at constant pressure and volume respectively, 
$p_0$ is the reference surface pressure and $R = c_p - c_v$ is the ideal gas constant. The velocity transport term is 
expressed in \eqref{eq::mom} in \emph{vector invariant} form as 
$(\nabla\times\boldsymbol{u})\times\boldsymbol{u} + \nabla(\boldsymbol{u}\cdot\boldsymbol{u})/2$. This form has the desirable
property that the rotational and potential components of the flow are treated as separate terms. For the appropriate choice 
of finite element spaces for these terms, we can ensure that there is no spurious projections between rotational and divergence
components of the flow in the discrete form \cite{Cotter and Shipton (2012)}.
As an alternative we may also express this term in \emph{advective} form as $\boldsymbol{u}\cdot\nabla\boldsymbol{u}$. We will 
investigate both forms in the proceeding section.

Unlike the shallow water system,
the compressible Euler equations contain an algebraic equation in the form of the ideal gas law \eqref{eq::eos}, such that 
the prognostic variables are given as $\boldsymbol{x}_h=(\boldsymbol{u}_h,\rho_h,\theta_h)$ and the diagnostic variable as 
$\boldsymbol{y}_h=\Pi_h$. Via 
\green{block factorisation of the coupled system (and approximate lumping of the non-diagonal velocity space mass matrix) }
one may derive a non-singular Helmholtz equation for the solution of $\Pi_h$ \cite{Maynard et. al. (2020)}, or alternatively the density weighted potential temperature 
$(\rho\theta)_h$\green{ if this is used as a prognostic variable instead of $\theta_h$ }
\cite{Lee (2021)}. Consequently any RoW method used to solve the above system as a Helmholtz 
problem, either directly or as a preconditioner, should also be applicable to index-2 PDAEs. 

One possible approximate Jacobian for the above system, which results in a Helmholtz problem for the Exner pressure
via repeated Schur complement decomposition \cite{Melvin et. al. (2019), Maynard et. al. (2020)} is given as
\begin{align}\label{eq::ce_W}
	\boldsymbol{\mathsf{W}}_{ce} &=
	\begin{bmatrix}\boldsymbol{\mathsf{C}} & \boldsymbol{\mathsf{0}} &
		\boldsymbol{\mathsf{P}}^{\Pi *}_{u\theta} & -\boldsymbol{\mathsf{G}}^{\theta *} \\
		\boldsymbol{\mathsf{D}}^{\rho *} & \boldsymbol{\mathsf{0}} &
		\boldsymbol{\mathsf{0}} & \boldsymbol{\mathsf{0}} \\
		\boldsymbol{\mathsf{P}}^{\theta *}_{\theta u} &
		\boldsymbol{\mathsf{0}} & \boldsymbol{\mathsf{0}} & \boldsymbol{\mathsf{0}} \\
		\boldsymbol{\mathsf{0}} & \boldsymbol{\mathsf{N}}_{\Pi}^{\rho *} & 
		\boldsymbol{\mathsf{P}}_{\Pi\theta}^{\theta *} & \boldsymbol{\mathsf{N}}_{\Pi}^{\Pi *} \\
	\end{bmatrix}\notag\\
	&\approx
	\begin{bmatrix}\frac{\partial F_{u}^h}{\partial\boldsymbol{u}^h} & \boldsymbol{\mathsf{0}} &
		\frac{\partial F_{u}^h}{\partial\theta^h} & \frac{\partial F_{u}^h}{\partial\Pi^h} \\
		\frac{\partial F_{\rho}^h}{\partial\boldsymbol{u}^h} & \boldsymbol{\mathsf{0}} &
		\boldsymbol{\mathsf{0}} & \boldsymbol{\mathsf{0}} \\
		\frac{\partial F_{\theta}^h}{\partial\boldsymbol{u}^h} &
		\boldsymbol{\mathsf{0}} & \boldsymbol{\mathsf{0}} & \boldsymbol{\mathsf{0}} \\
		\boldsymbol{\mathsf{0}} & \frac{\partial G_{\Pi}^h}{\partial\rho^h} & 
		\frac{\partial G_{\Pi^h}}{\partial\theta^h} & \frac{\partial G_{\Pi}^h}{\partial\Pi^h} \\
	\end{bmatrix},
\end{align}
where $\rho*$, $\theta*$ and $\Pi*$ are reference profiles for the density, potential temperature and Exner
pressure respectively, derived from the prognostic variables at the time of the Jacobian assembly. For a full 
description of these operators in the context of a $H(div)$ conforming finite element discretisation with 
$\boldsymbol{u}_h\in\mathbb{W}_2\subset H(div)$, $\rho_h,\Pi_h\in\mathbb{W}_3\subset L^2$,
$\theta_h\in\mathbb{W}_{cp}\subset H(div)$ (where $\mathbb{W}_{cp}$ is the set of bases in the subspace of 
$H(div)$ consisting of scalar functions that are $C^0$ continuous in the vertical dimension only) see 
\cite{Melvin et. al. (2019)}.

\subsubsection{Implementation of transport terms in \emph{LFRic}}

In order to allow for longer time steps in LFRic with a fixed number of iterations, the mass and 
temperature transport terms in the right hand sides of \eqref{eq::ce} (and the velocity when in advective form)
are evaluated explicitly \green{at each nonlinear iteration } over a series of $M$ smaller CFL dependent sub-steps using a
transport velocity $\boldsymbol{u}^t_h$.
Expressing the generalised transport operator at iteration $i$ as $\mathcal{F}^{(i)}(\psi_h;\boldsymbol{u}^t_h)$ for 
$\rho_h,\theta_h,\boldsymbol{u}_h\in\psi_h$, 
where for $\psi_h=\rho_h$ $\mathcal{F}^{(i)}(\rho_h;\boldsymbol{u}_h)$ is the discrete form of the flux form transport operator 
$F_{\rho,h}$ \eqref{eq::mass}, and otherwise this is the material form transport operator \eqref{eq::temp},
the sub-stepped explicit transport is given as:
\begin{subequations}\label{eq::explicit_adv}
\begin{align}
	\Delta tF_{\psi,h} &= \Delta t\int_{t^n}^{t^{n+1}}\mathcal{F}^{(i)}(\psi^{n}_h;\boldsymbol{u}^t_h)\mathrm{d}t \notag \\ 
	&= \sum_{m=1}^M\psi_h^m - \psi_h^{m-1},\\
	\psi_h^m - \psi_h^{m-1} &= \frac{\Delta t}{M}\sum_p^Pb_p\mathcal{F}^{(i)}(\hat{\psi}_h^{p,m};\boldsymbol{u}^t_h),\\
	\hat{\psi}_h^{p,m} &= \psi_h^{m-1} + \sum_{q=1}^{p-1}a_{pq}\mathcal{F}^{(i)}(\hat{\psi}_h^{q,m};\boldsymbol{u}^t_h).
\end{align}
\end{subequations}
\red{where $a_{pq}$, $b_p$ are the coefficients and weights respectively for the explicit Runge-Kutta substep $m$}.
For the CN scheme a time centered velocity is used for the transport as 
$\boldsymbol{u}^t_h = (\boldsymbol{u}_h^n + \boldsymbol{u}_h^{(i-1)})/2$.
For the RoW scheme we cannot use a time centered velocity, owing to the specific reconstruction of the 
state at iteration $i$ as given in \eqref{eq::ros_x_star}. As such there are two possible formulations
for the explicit transport in the context of the RoW scheme. The first is to avoid sub-stepping, and 
just use a single forward Euler step at the \emph{current} state as
$F_{\psi,h} = \mathcal{F}^{(i)}(\psi_*^{(i)};\boldsymbol{u}_*^{(i)})$. This approach is consistent with 
the RoW method \eqref{eq::ros}, but is limited to time steps with small advective CFL numbers due 
to the absence of sub-stepping.

The second approach is to integrate the \emph{initial} stage $\psi^n_h$ over the time level using the 
incremental solution for the velocity at RoW stage $i$, $\boldsymbol{k}_u^{(i)}$ as:
\begin{subequations}\label{eq::row_adv}
\begin{align}
	F_{\psi,h}^{(1)} = &\int_{t=t^n}^{t^{n+1}}\mathcal{F}({\psi}_h^n;\boldsymbol{u}^n)\mathrm{d}t\\
	F_{\psi,h}^{(2)} = &F_{\psi,h}^{(1)} + \alpha_{11}\mathcal{F}({\psi}_h^n;\boldsymbol{k}_u^{(1)})\\
	F_{\psi,h}^{(3)} = &F_{\psi,h}^{(1)} + \alpha_{21}\mathcal{F}({\psi}_h^n;\boldsymbol{k}_u^{(1)}) + \alpha_{22}\mathcal{F}({\psi}_h^n;\boldsymbol{k}_u^{(2)})\\
	F_{\psi,h}^{(4)} = &F_{\psi,h}^{(1)} + \alpha_{31}\mathcal{F}({\psi}_h^n;\boldsymbol{k}_u^{(1)}) + \alpha_{32}\mathcal{F}({\psi}_h^n;\boldsymbol{k}_u^{(2)}) 
	\notag\\
	&+\int_{t=t^n}^{t^{n+1}}\mathcal{F}({\psi}_h^n;\boldsymbol{k}_u^{(3)})\mathrm{d}t.
\end{align}
\end{subequations}
The first stage, $i=1$ is integrated in the same fashion as the CN scheme \eqref{eq::explicit_adv}.
If the velocity increment is small with respect to the full velocity from which the CFL is determined, then we may
update the subsequent stages using only single forward Euler steps, which are cheap to compute. There is no
requirement on the last stage, $i=4$, that the increment here be computed using forward Euler, and so this
increment may once again be sub-stepped if required.

\green{The transported quantities used to compute the fluxes, $\psi_h$,
may be derived from the values in neighbouring cells using either a centered or an upwinded reconstruction
(which smoothes out oscillations via the dissipation of higher moments such as energy). We will explore the 
implications of both centered and upwinded fluxes in Sections \ref{sec::bw_stab} and \ref{sec::bw_time} 
respectively.}

\section{Results}

\subsection{Rotating shallow water: shear flow instability on the sphere}\label{sec::galewsky}

We compare the results of four iterations of the Crank-Nicolson discretisation (CN4) to a variety of four stage
RoW schemes in terms of both stable time step and energetic profiles for a standard shear flow
instability test case on the sphere \cite{Galewsky et. al. (2004)}, run for 12 days so as to ensure that the schemes remain
stable with the specified time step for a mature turbulent state. \red{No off-centering is applied to the CN4 scheme,
as is often done to increase stability at the expense of a reduced order of accuracy, however this is explored for the 
3D compressible Euler equations in the proceeding section. } The various RoW schemes are detailed in Table I.
\begin{table*}[t]
\begin{center}
\begin{tabular}{|c|c|c|c|}
	\hline
	Scheme & Reference & Index & Stability \\
	\hline
	ROS34PW2  & \cite{Rang and Angermann (2005)}   & 1 & $R(\infty)=0$ \\
	ROS34PW3  & \cite{Rang and Angermann (2005)}   & 1 & $R(\infty)\approx 0.63$ \\
	ROSI2\blue{P}w    & \cite{Rang and Angermann (2008)}   & 2 & $R(\infty)=0$, $\boldsymbol{\mathsf{W}}=\frac{\delta F}{\delta\boldsymbol{x}} + \mathcal{O}(\Delta t)$ \\
	ROSI2\blue{P}W    & \cite{Rang and Angermann (2008)}   & 2 & $R(\infty)=0$ \\
	ROS34PRW  & \cite{Rang (2013)}   & 2 & $R(\infty)=0$ \\
	ROS3PRL2  & \cite{Rang (2015), Lang and Teleaga (2007)} &   & $R(\infty)=0$ \\
	ROWDAIND2 & \cite{Lubich (1990)} & 2 & $R(\infty)=0$ \\
	\hline
\end{tabular}
\end{center}
	\caption{Rosenbrock-Wanner schemes used in this study. \red{Index 2 schemes are consistent for the 
	Helmholtz Exner pressure preconditioner used in the \emph{LFRic} model, and schemes with $R(\infty)=0$ 
	are L-stable, with the amplification factor approaching zero for the largest eigenvalues.}}
\end{table*}

In each case the same $H(div)$ conforming finite element method is applied for the spatial 
discretisation \cite{Lee and Palha (2018), Lee et. al. (2022)}, using $32\times 32$ third order finite elements on each panel 
of the cubed sphere \green{for an average grid spacing of $\Delta x\approx 96{km}$}. 
The non-linear potential enstrophy cascade to grid scales is stabilised via the 
anticipated potential vorticity method \cite{Sadourny and Basdevant (1985)} with an upwinding parameter of $\Delta t/2$,
\blue{and energy is exactly conserved by the spatial discretisation}.
\green{The mixed finite element discretisation used here has the same compatibility relations as in the \emph{LFRic}
model \cite{Melvin et. al. (2019)}, but with polynomials two degrees higher than those used to solve the 3D compressible
Euler equations in the proceeding section. The shallow water model also differs in its representation of the 
transport terms, which like all other terms are evaluated instantaneously as per \eqref{eq::ros}, and not 
explicitly as in \eqref{eq::explicit_adv}. } \green{The shallow water model uses 6 point Gauss-Lobatto-Legendre
quadrature. While this is exact for $9^{th}$ order polynomials, the Jacobian transformation used to represent the
surface of the sphere uses trigonometric functions, which cannot be exactly integrated at any order. Consequently
there will be some small aliasing errors present, which may also affect the solver convergence and stability.}

\begin{figure}[!hbtp]
\begin{center}
\includegraphics[width=0.48\textwidth,height=0.36\textwidth]{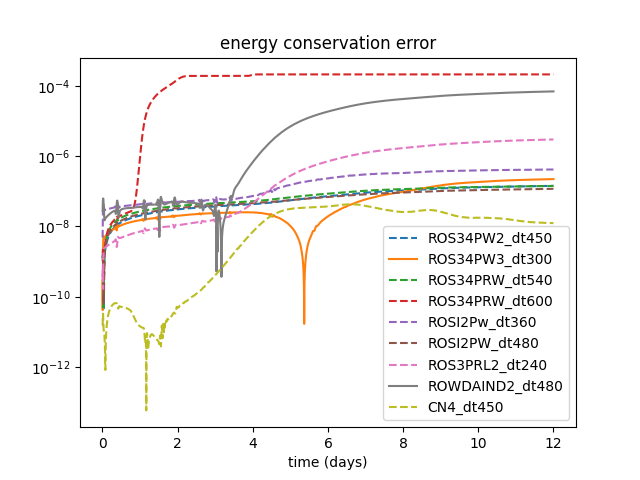}
\includegraphics[width=0.48\textwidth,height=0.36\textwidth]{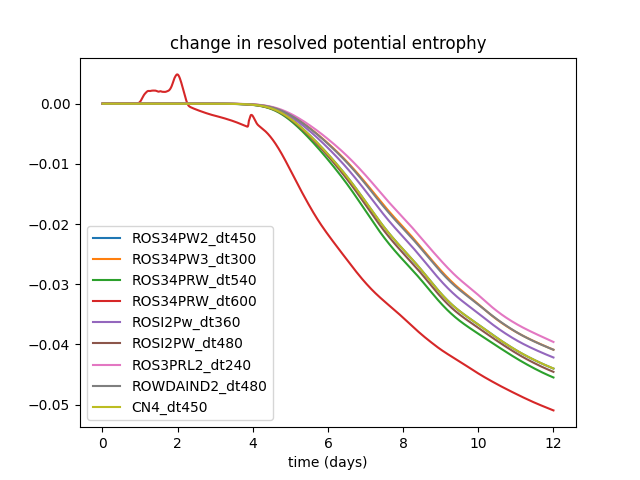}
\caption{Normalised energy (top) and potential enstrophy (bottom) conservation errors for the 
different integrators for the shear flow instability test case on the sphere over 12 days at the
maximum observable stable time step for each scheme. Dashed lines for the energy conservation error 
indicate energy decay and solid lines indicate growth.}
\label{fig::sw_1}
\end{center}
\end{figure}

Figure \ref{fig::sw_1} shows the normalised energy and potential enstrophy conservation errors for the 
different four-stage RoW and the CN4 scheme for the shear flow instability test case on 
the sphere. These are computed by globally integrating the total energy $E_{sw}$ and potential enstrophy 
$Z_{sw}$ over the domain $\Omega$ as
\begin{subequations}
	\begin{align}
		E_{sw} &= \int\frac{1}{2}h_h\boldsymbol{u}_h\cdot\boldsymbol{u}_h + \frac{g}{2}h_h^2\mathrm{d}\Omega\\
		Z_{sw} &= \int\frac{1}{2}h_hq_h^2\mathrm{d}\Omega,
	\end{align}
\end{subequations}
where \red{the discrete form of the potential vorticity, $q_h$, can be found in \cite{Bauer et. al. (2018), Lee et. al. (2022)}}.

For the energy conservation error plot the solid lines \blue{corresponding to the ROS34PW3 and ROWDAIND2 schemes } 
indicate a long time energy growth, so that the solution will ultimately become unstable. \blue{For the rest of the 
schemes }, dashed lines indicate long time energy decay and hence stability. Of all the schemes, one in particular
allows for stable simulation using significantly longer time steps, ROS34PRW 
\cite{Rang (2013)}. Some care must be taken however, as while this scheme is stable for time steps
of up to 600 seconds, at these long times
the shear flow instability occurs at the incorrect wave-number,
\red{perhaps due to dispersion errors in the representation of the gravity waves which trigger the instability}.
This is also observed in the potential enstrophy conservation error, which exhibits an anomalous
bump as the dynamics transition to the incorrect wave number. This problem is not observed and 
the dynamics evolve correctly for a time step of 540 seconds for the ROS34PRW scheme, 
which is still $20\%$ longer than the maximum stable time step of the CN4 scheme at 450 seconds.
The potential enstrophy conservation error is broadly representative of the richness of the turbulence
present in the solution. These errors broadly correlate with the length of the time steps for the 
different schemes, with the schemes with smaller stable time steps, such as ROS3PRL2 exhibiting the
smallest potential enstrophy conservation error and the ROS34PRW the greatest.

\begin{figure}[!hbtp]
\begin{center}
\includegraphics[width=0.48\textwidth,height=0.36\textwidth]{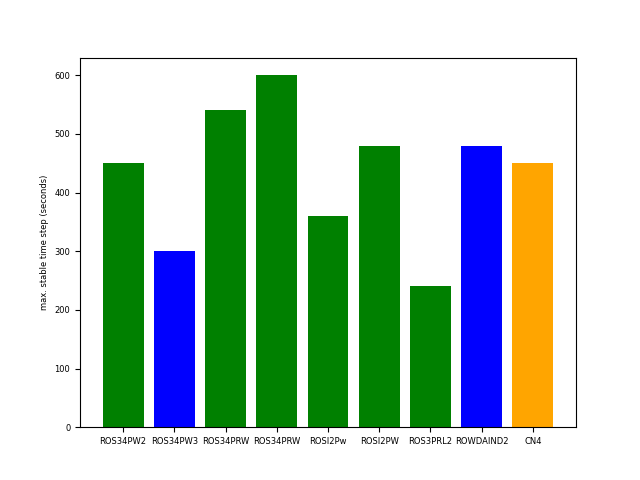}
\caption{Maximum stable time steps for the different RoW and CN
	schemes over 12 days of the shear flow instability test case on the sphere.}
\label{fig::sw_2}
\end{center}
\end{figure}

The maximum stable time steps are more clearly observed in the bar chart in Fig. \ref{fig::sw_2}. 
Here the schemes that exhibit positive energy error growth are given in blue, while the schemes 
that are long time stable are in green. The CN4 scheme is given as a reference in orange. The CN4
scheme is somewhere in the middle in terms of maximum stable time step, with several 
RoW schemes allowing for longer time steps. The ROS34PRW scheme is presented twice, 
once for its maximum stable time step of 600 seconds, and once for its
maximum physically correct time step of 540 seconds.

\red{The maximum stable time step was also investigated for the CN scheme at convergence.
This was found to be the same as for four iterations (450 seconds), suggesting that the nonlinear errors
after four iterations are well resolved with respect to other sources of error, such as approximations made
in the form of the Jacobian and inexact spatial integration as mentioned above.}

\subsection{3D compressible Euler: baroclinic instability on the sphere}\label{sec::bw}

The RoW integrators are compared against the CN scheme for the 3D compressible Euler equations
at planetary scales using a standard test case for a baroclinic wave triggered by a velocity
perturbation in an otherwise geostrophically and hydrostatically balanced atmosphere \cite{Ullrich et. al. (2014)}
on z-levels \red{as used in the LFRic model. }
We compare the schemes in two different contexts, firstly in terms of stability and energetics for a 
centered reconstruction of the fluxes in the transport terms, which minimises the amount of internal 
dissipation within the model in order to highlight differences in the representation of dynamics between 
the schemes.
\blue{Secondly we compare the schemes in terms of computational performance using upwinded fluxes
in order to damp internal oscillations and allow for longer time steps.}

\subsubsection{Stability with centered transport terms}\label{sec::bw_stab}

The schemes are compared at two different spatial/temporal resolutions C48;
$6\times48\times 48$ lowest order elements ($\Delta x\approx 192km$, $\Delta t=1800s$) and C96;
$6\times96\times 96$ lowest order elements ($\Delta x\approx 96km$, $\Delta t=900s$), both using 
30 vertical levels. \blue{In the context of the mixed finite element spatial discretisation used here, lowest order
elements refer to piecewise constant Exner pressure and density in each element, velocities that are
piecewise linear in the normal direction and piecewise constant in the tangent directions, and a 
potential temperature that is piecewise linear in the vertical direction and piecewise constant
in the horizontal directions \cite{Melvin et. al. (2019)}. } No external dissipation
was used in these simulations, so in all cases the solution ultimately becomes unstable. 
\green{By avoiding the use of external damping terms we are better able to compare the 
internal stability of the different time integration schemes. }
In order to maintain the stability of the CN4 scheme, this was run with the potential temperature 
transport term in the Jacobian, $\boldsymbol{\mathsf{P}}^{\theta *}_{\theta u}$ in \eqref{eq::ce_W}, 
\red{damped } in favor of the future time level, such that the factor of $\gamma=1$ in \eqref{eq::cn}
for this term only (and is kept as $\gamma=1/2$ for all other terms). 
In all cases a simple forward Euler integration of the transport terms \eqref{eq::explicit_adv} 
was applied for a single sub-step at each iteration (with this being evaluated at the current time 
level for the RoW schemes). This ensures that the transport terms are evaluated instantaneously 
so as to be faithful to the original CN and RoW formulations.

The globally integrated kinetic (horizontal and vertical), potential and internal energies are computed respectively over the 
full domain $\Omega$ at each time step as 
\begin{subequations}
\begin{align}
	K_h &= \int\frac{1}{2}\rho_h\boldsymbol{v}_h\cdot\boldsymbol{v}_h\mathrm{d}\Omega \\
	K_v &= \int\frac{1}{2}\rho_h w_h^2\mathrm{d}\Omega \\
	P &= \int gz\rho_h\mathrm{d}\Omega \\
	I &= \int c_v\rho_h\theta_h\Pi_h\mathrm{d}\Omega,
\end{align}
\end{subequations}
where $\boldsymbol{v}_h$ and $w_h$ are the horizontal and vertical velocities respectively, 
\green{and gravity, $g$, is assumed to be constant. The kinetic energy is separated into its
horizontal and vertical components. This makes the evolution of the vertical component,
which is significantly smaller than the horizontal component in most atmospheric configurations,
more visible, and helps to illustrate differences in the evolution of the baroclinic instability
between the different schemes.}

\begin{figure}[!hbtp]
\begin{center}
\includegraphics[width=0.48\textwidth,height=0.36\textwidth]{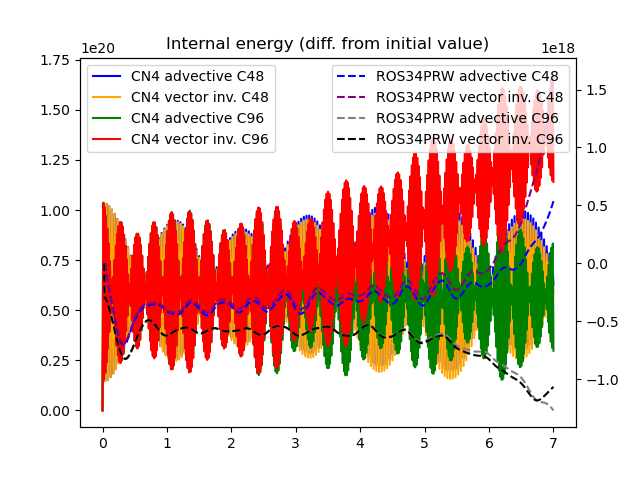}
\includegraphics[width=0.48\textwidth,height=0.36\textwidth]{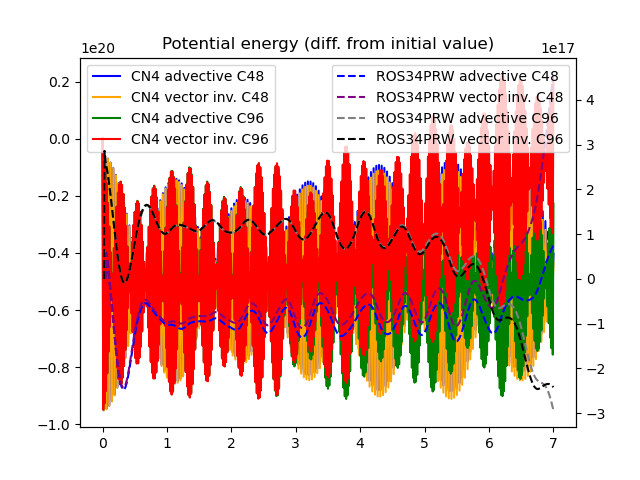}
	\caption{Internal (top) and potential (bottom) energy evolution for the baroclinic
	wave test case for the CN4 (advective and vector invariant) and ROS34PRW (vector invariant) 
	schemes at the C48 and C96 resolutions. Note the different scales on the vertical axes for
	the CN and RoW schemes.}
\label{fig::ce_bw_1}
\end{center}
\end{figure}

\begin{figure}[!hbtp]
\begin{center}
\includegraphics[width=0.48\textwidth,height=0.36\textwidth]{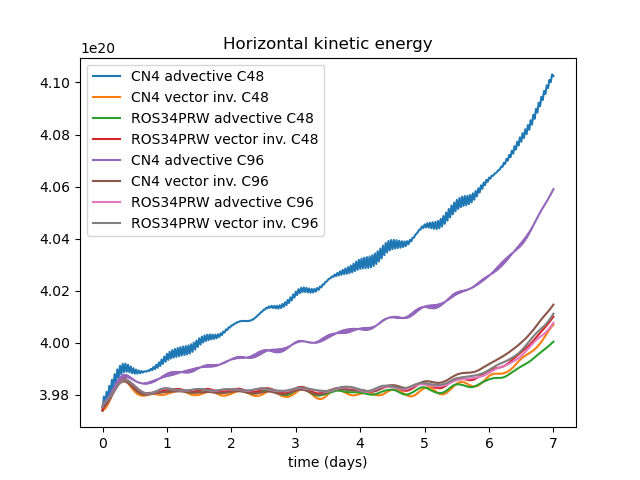}
\includegraphics[width=0.48\textwidth,height=0.36\textwidth]{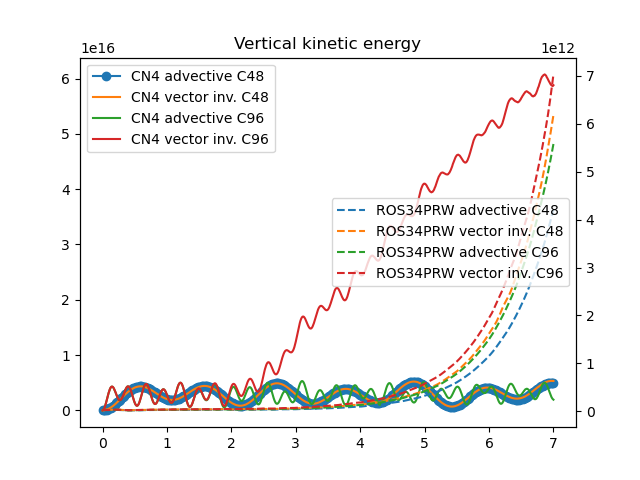}
	\caption{Horizontal (top) and vertical (bottom) kinetic energy evolution 
	for the baroclinic wave test case for the CN4 (advective and vector invariant) 
	and ROS34PRW (vector invariant) schemes at the C48 and C96 resolutions. Note the 
	different scales on the vertical axes for the CN and RoW schemes for the 
	vertical kinetic energy.}
\label{fig::ce_bw_2}
\end{center}
\end{figure}

As observed in the internal and potential energy evolution as shown in Fig. \ref{fig::ce_bw_1} the CN4
schemes exhibit a spurious oscillation on a time scale of $2\Delta t$, that is not present for the 
ROS34PRW scheme, which show a much smaller oscillation on a time scale of approximately 12 hours
independent of time step size. This oscillation is consistent with the temporal oscillation observed
in the horizontal kinetic energy in Fig. \ref{fig::ce_bw_2}, \red{and is perhaps a result of inexact
hydrostatic balance in the initial condition that is quickly damped by the L-stable RoW scheme, but not
by the CN4 scheme}. The growth of the baroclinic instability
is observed in the evolution of the vertical kinetic energy for the ROS34PRW scheme also in 
Fig. \ref{fig::ce_bw_2}. This result is consistent with previous observations using a high order mixed
finite element model with horizontally explicit/vertically implicit time stepping and exact energy 
conservation for the implicit vertical solve \cite{Lee (2021)}. However for the CN4 scheme, this signal is 
insignificant with respect to the vertical kinetic energy signal associated with the internal-potential 
buoyancy oscillation \red{(left axis)}, which at $\mathcal{O}(10^{15})$ Joules is approximately $100$ times greater than 
the vertical kinetic energy associated with the baroclinic instability, \red{as observed for the ROS34PRW scheme on the right axis}. 

This persistence of oscillation is perhaps a consequence of the neutral stability of the CN4 temporal scheme with 
$\gamma=1/2$. Using the L-stable $\gamma$ values described for the four iteration CN
scheme in Section \ref{sec::1.1}, this oscillation is suppressed, however the simulation is rapidly observed 
to be unstable (and so is not presented here). An alternative, stable method for suppressing this 
oscillation is to off-center the CN scheme in favor of the new time level (by a factor 
of $0.55$), however this leads to a degradation of accuracy.
Results using this off-centering for the CN4 scheme will be presented below.

In terms of the internal and potential energy evolution, only the ROS34PRW scheme at the C96 resolution
gives observably consistent results with respect to those previously published using an exact energy conserving vertical integrator 
\cite{Lee (2021)}, where both the internal and potential energy trend downward with time in order to balance the
growth in kinetic energy due to the baroclinic instability.

\begin{figure}[!hbtp]
\begin{center}
\includegraphics[width=0.48\textwidth,height=0.36\textwidth]{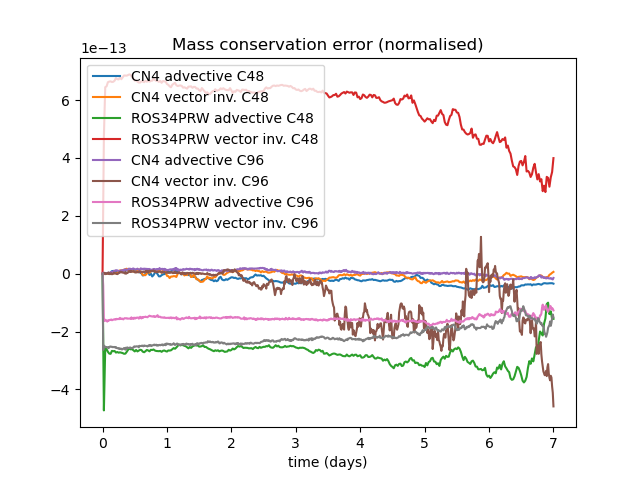}
\includegraphics[width=0.48\textwidth,height=0.36\textwidth]{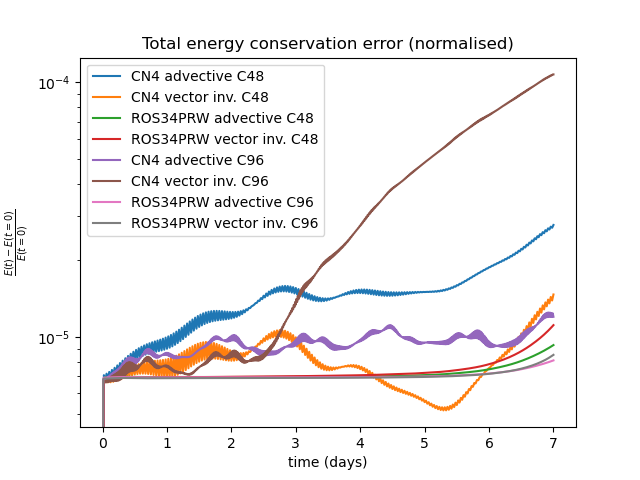}
	\caption{Normalised \blue{mass (top) } and total energy conservation error (bottom)
	for the baroclinic wave test case for the CN4 (advective and vector invariant) 
	and ROS34PRW (vector invariant) schemes at the C48 and C96 resolutions.}
\label{fig::ce_bw_3}
\end{center}
\end{figure}

\blue{Normalised mass conservation errors for the different configurations are presented in Fig. \ref{fig::ce_bw_3}.
Since in all cases the mass continuity equation represented by a flux form finite volume scheme, the
conservation errors are all 
close to machine precision. However the RoW configurations manifest errors an order of magnitude
greater than those for the CN scheme. }
The total energy conservation error is also given in Fig. \ref{fig::ce_bw_3}, where it is observed 
that the CN4 scheme has significantly greater total energy growth than the ROS34PRW scheme at the
same resolution. In particular, the vector invariant form of the CN4 integrator at C96 resolution exhibits 
a rapid growth of the energy conservation error that suggests the onset of numerical instability.
This is reflected in the fact that this configuration requires two forward Euler sub-steps for the 
transport terms at each nonlinear iteration, whereas each of the other configurations requires
just a single forward Euler sub-step for the transport \blue{due to the moderate velocity 
magnitudes in comparison to the CFL number in this test case. } 
\red{The instability observed for the vector invariant form of the CN4 discretisation is perhaps 
due to this configuration being closer to a neutrally stable state than the more
overtly damped advective form (or the L-stable vector invariant ROS34PRW scheme), and so is more 
susceptible to instability as potentially triggered by the high frequency temporal oscillation 
observed for the CN4 scheme in Fig. \ref{fig::ce_bw_1}.}

\green{While the energetic results presented here are revealing in terms of how the different
time stepping schemes effect the internal dynamics, we note that these errors may be small compared
to those introduced by external forcing terms such as parameterisations of physical processes and 
the coupling to other model components.}

\begin{figure}[!hbtp]
\begin{center}
\includegraphics[width=0.48\textwidth,height=0.36\textwidth]{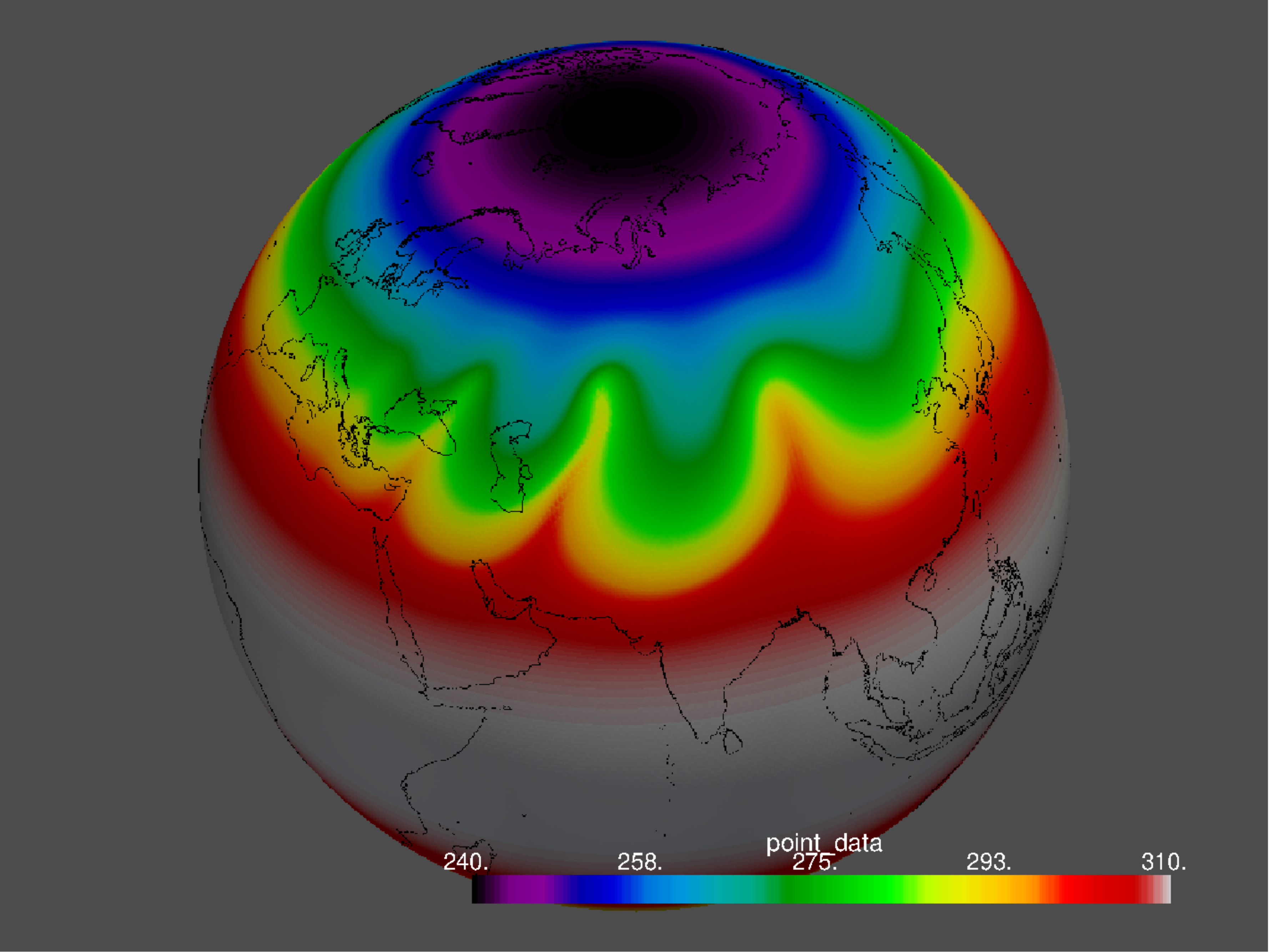}
\includegraphics[width=0.48\textwidth,height=0.36\textwidth]{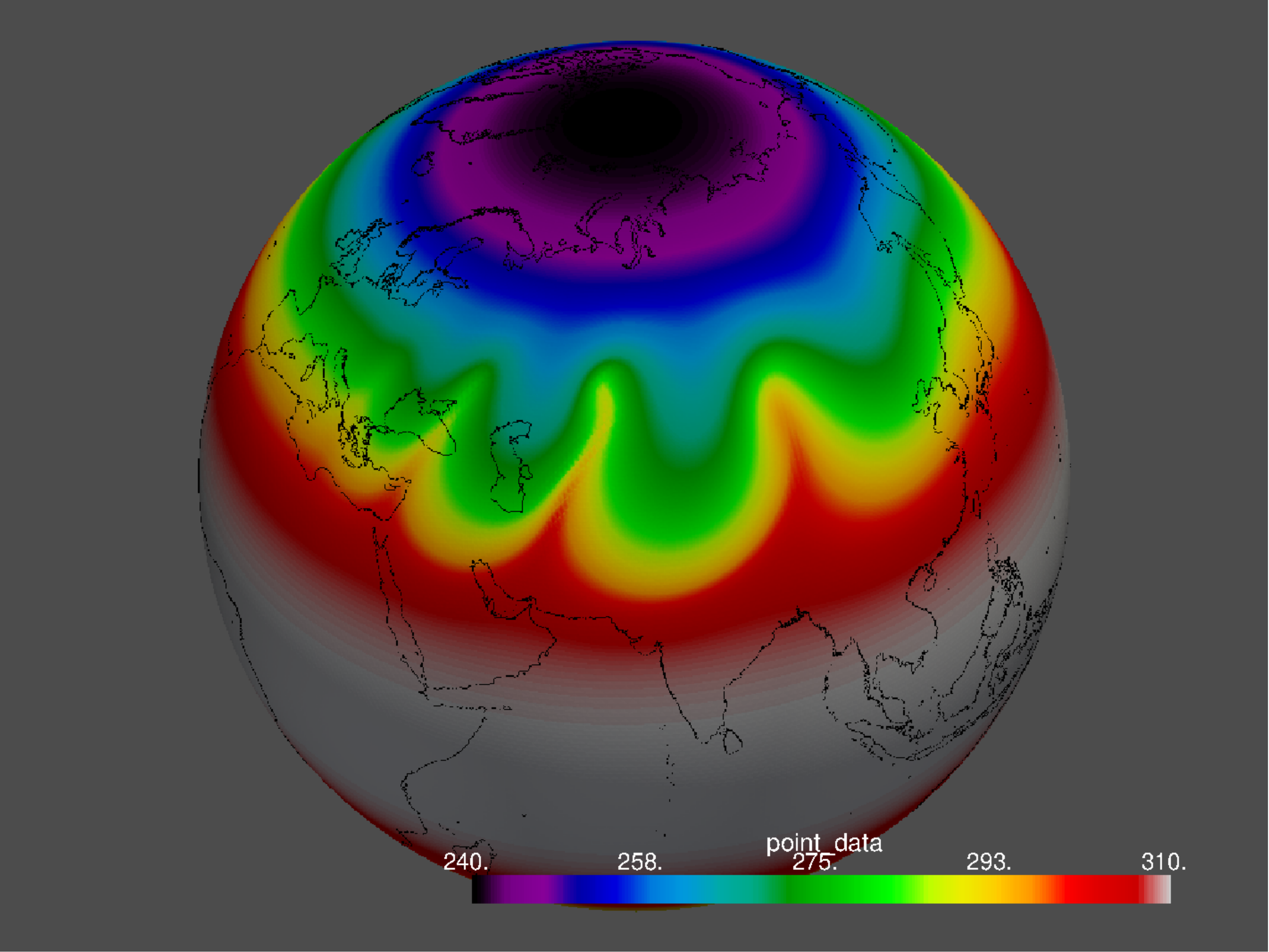}
	\caption{Surface potential temperature for the CN4 vector invariant (top) and advective form (bottom) 
	at day 7 for the baroclinic wave test case at the C96 resolution.}
\label{fig::ce_bw_4_1}
\end{center}
\end{figure}

\begin{figure}[!hbtp]
\begin{center}
\includegraphics[width=0.48\textwidth,height=0.36\textwidth]{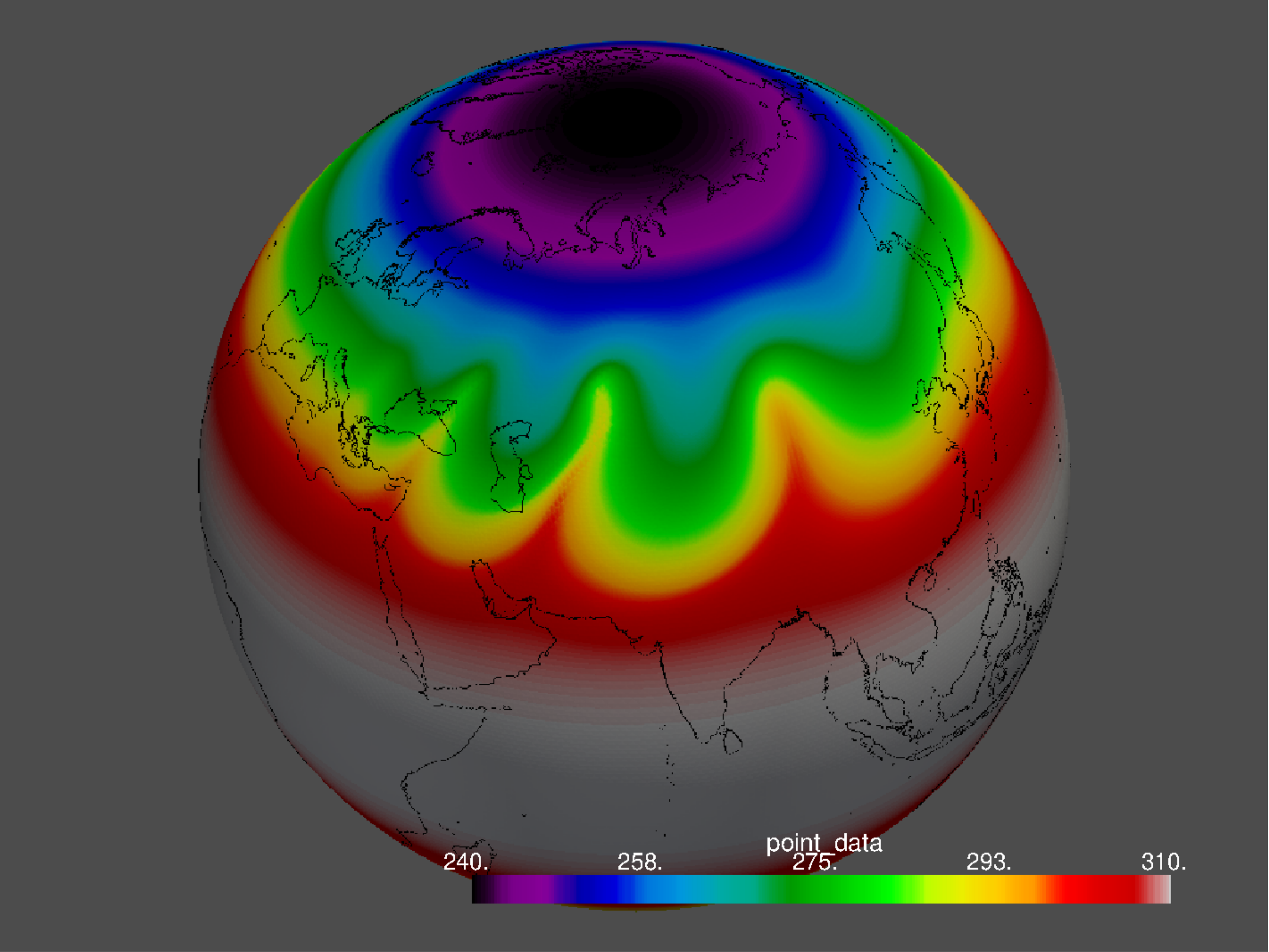}
\includegraphics[width=0.48\textwidth,height=0.36\textwidth]{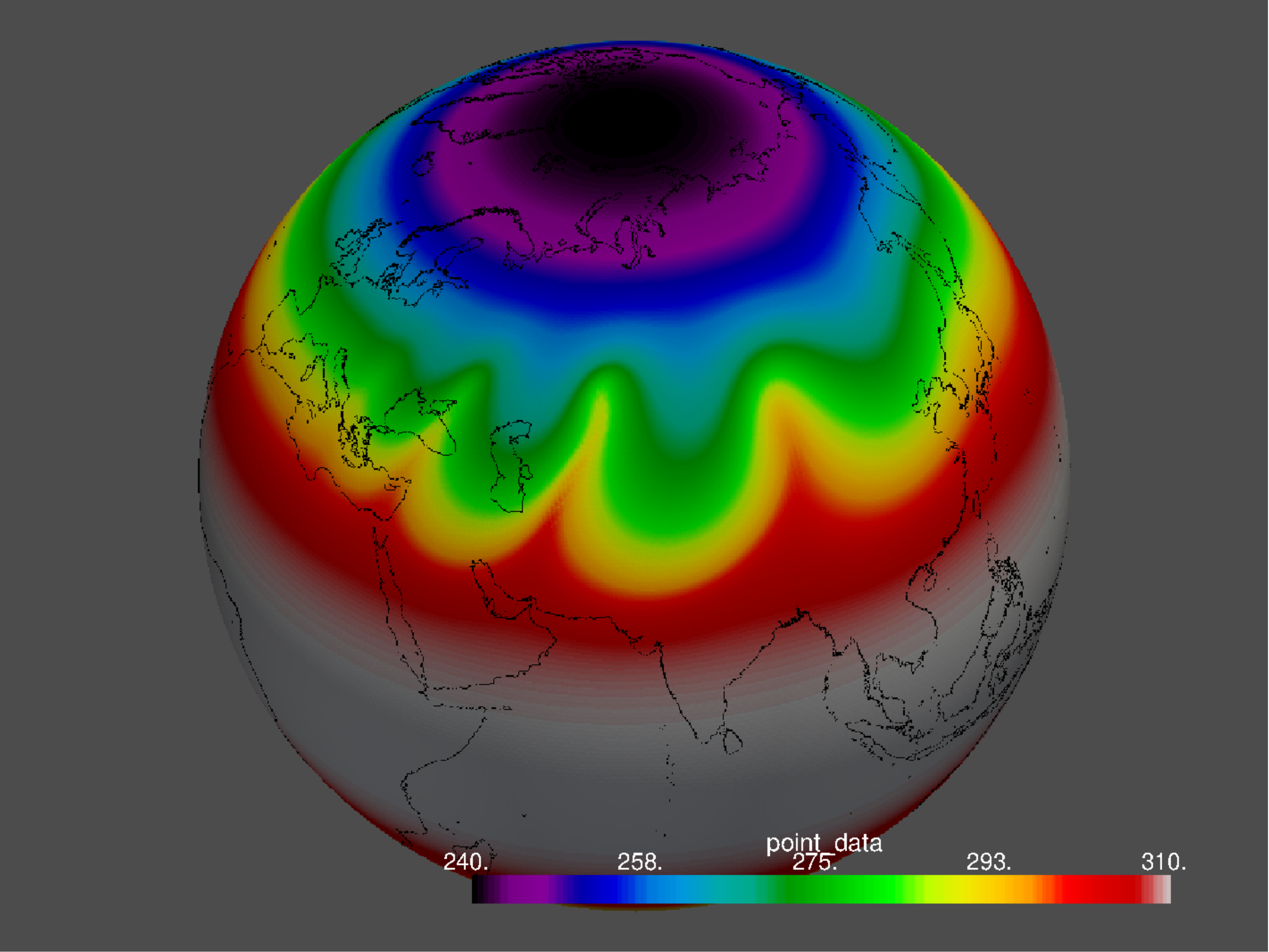}
	\caption{Surface potential temperature for the ROS34PRW vector invariant (top) and advective form (bottom) 
	at day 7 for the baroclinic wave test case at the C96 resolution.}
\label{fig::ce_bw_4_2}
\end{center}
\end{figure}

\begin{figure}[!hbtp]
\begin{center}
\includegraphics[width=0.48\textwidth,height=0.36\textwidth]{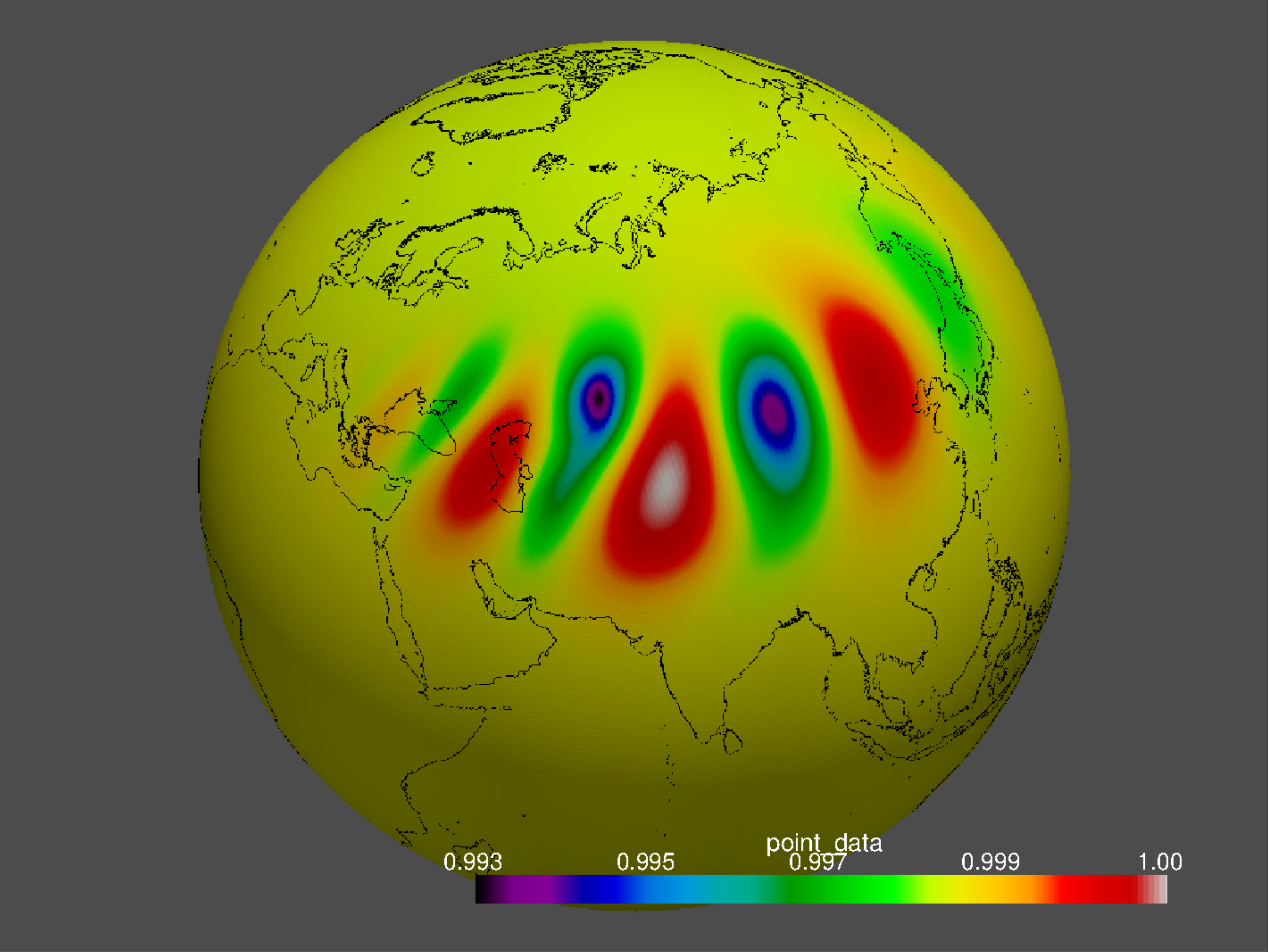}
\includegraphics[width=0.48\textwidth,height=0.36\textwidth]{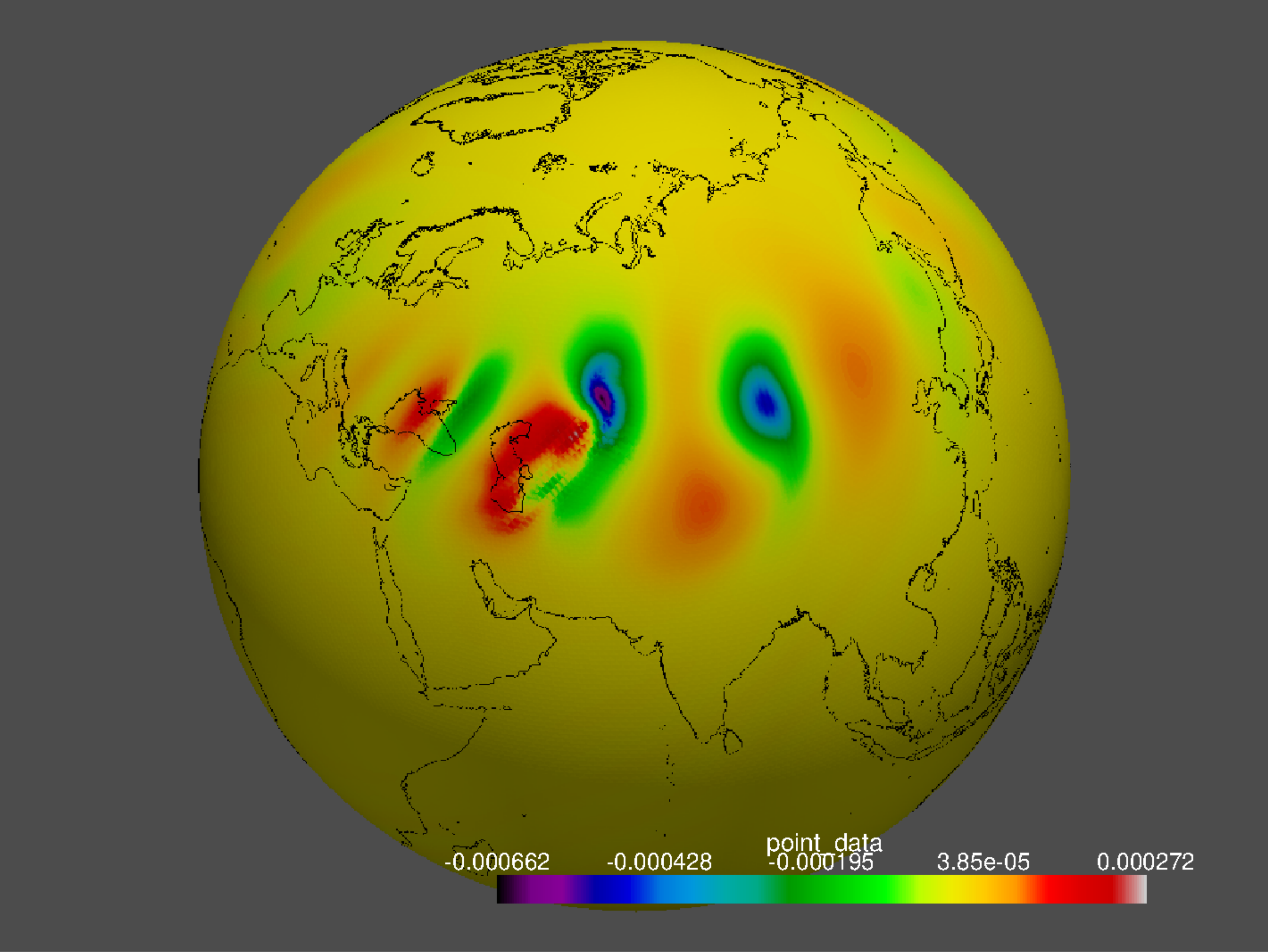}
	\caption{Surface level Exner pressure for the ROS34PRW vector invariant form (top) and 
	normalised difference with respect to the advective form (bottom) 
	at day 7 for the baroclinic wave test case at the C96 resolution.}
\label{fig::ce_bw_5_2}
\end{center}
\end{figure}

\begin{figure}[!hbtp]
\begin{center}
\includegraphics[width=0.48\textwidth,height=0.36\textwidth]{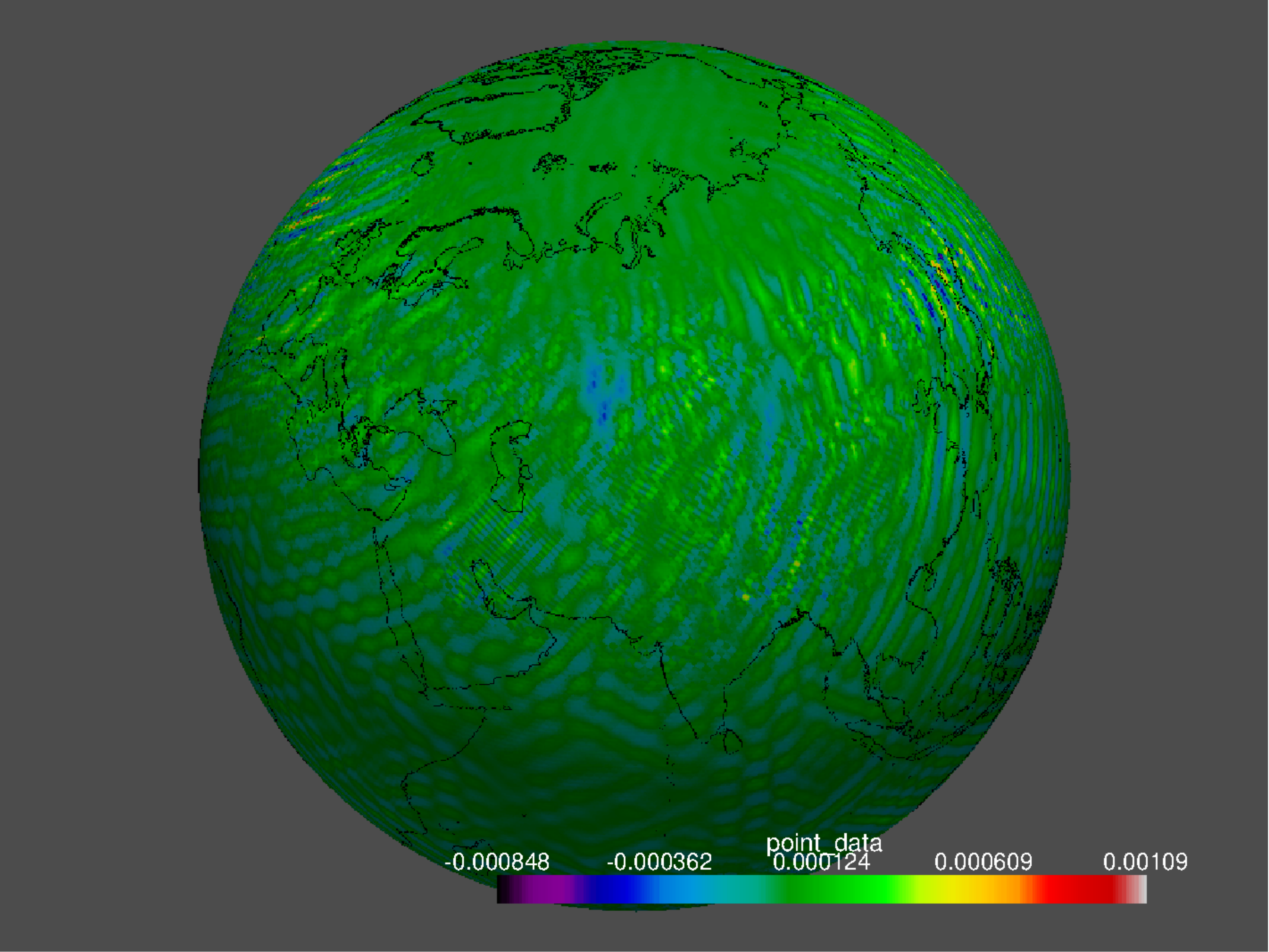}
\includegraphics[width=0.48\textwidth,height=0.36\textwidth]{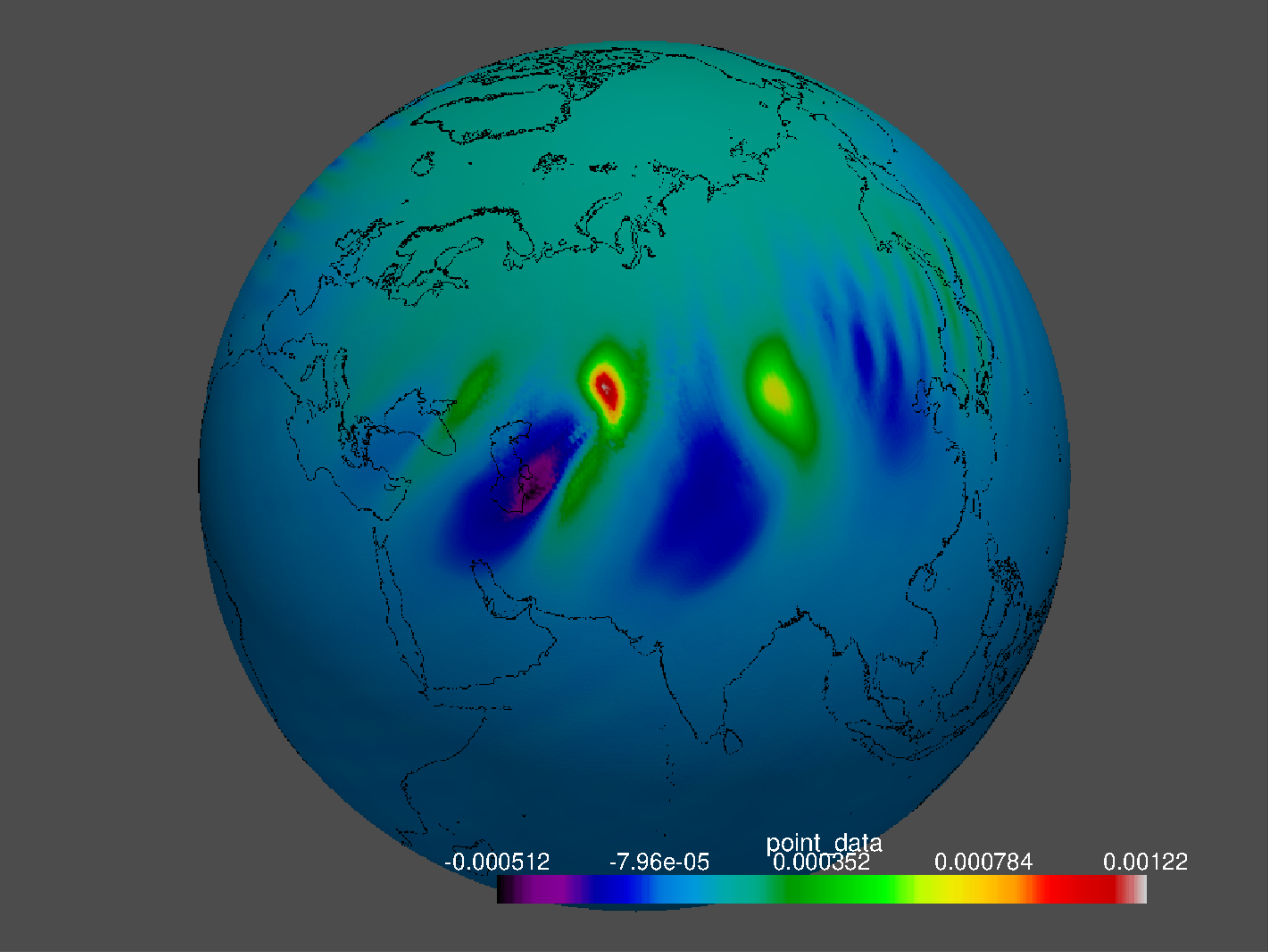}
	\caption{Surface level Exner pressure normalised difference of the 
	CN4 vector invariant (top) and advective form (bottom) with respect to the 
	ROS34PRW vector invariant form
	at day 7 for the baroclinic wave test case at the C96 resolution.}
\label{fig::ce_bw_5_1}
\end{center}
\end{figure}

The lowest level potential temperature and Exner pressure are shown at day 7 for the C96 
resolution in Figs \ref{fig::ce_bw_4_1}, \ref{fig::ce_bw_4_2} and \ref{fig::ce_bw_5_1}, \ref{fig::ce_bw_5_2} 
respectively for the CN4 and ROS34PRW schemes. \red{For the Exner pressure these are presented as a normalised
difference with respect to the vector invariant ROS34PRW scheme. } While the results are in broad agreement 
for the potential temperature, there is a spurious meridional variation in surface 
pressure observed for the advective CN4, \red{however this variation is small with respect to
differences in the solution for the baroclinic wave}. Additionally, 
spurious small scale oscillations are observed for the CN4 vector invariant scheme. 

\begin{figure}[!hbtp]
\begin{center}
\includegraphics[width=0.48\textwidth,height=0.36\textwidth]{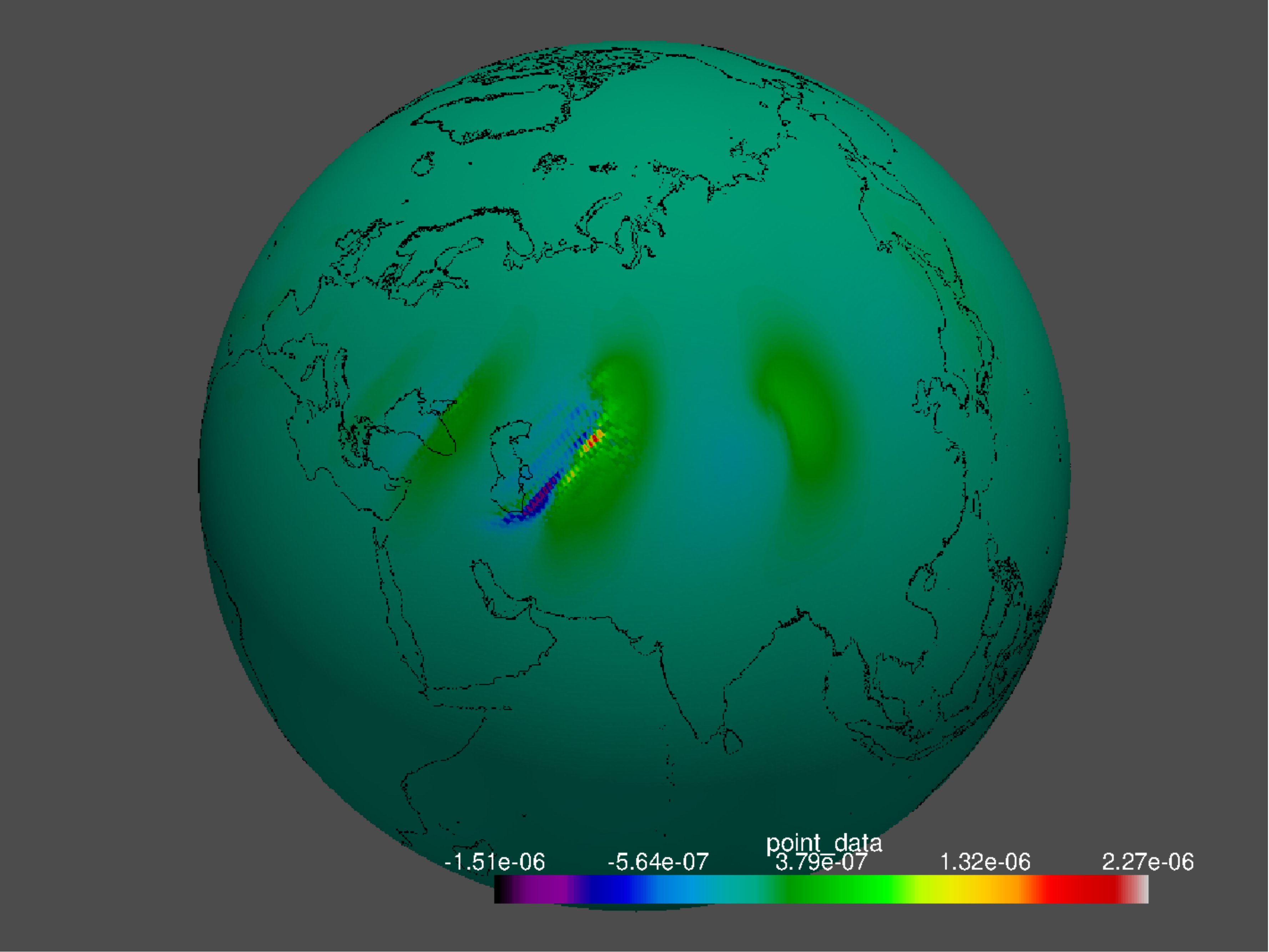}
\includegraphics[width=0.48\textwidth,height=0.36\textwidth]{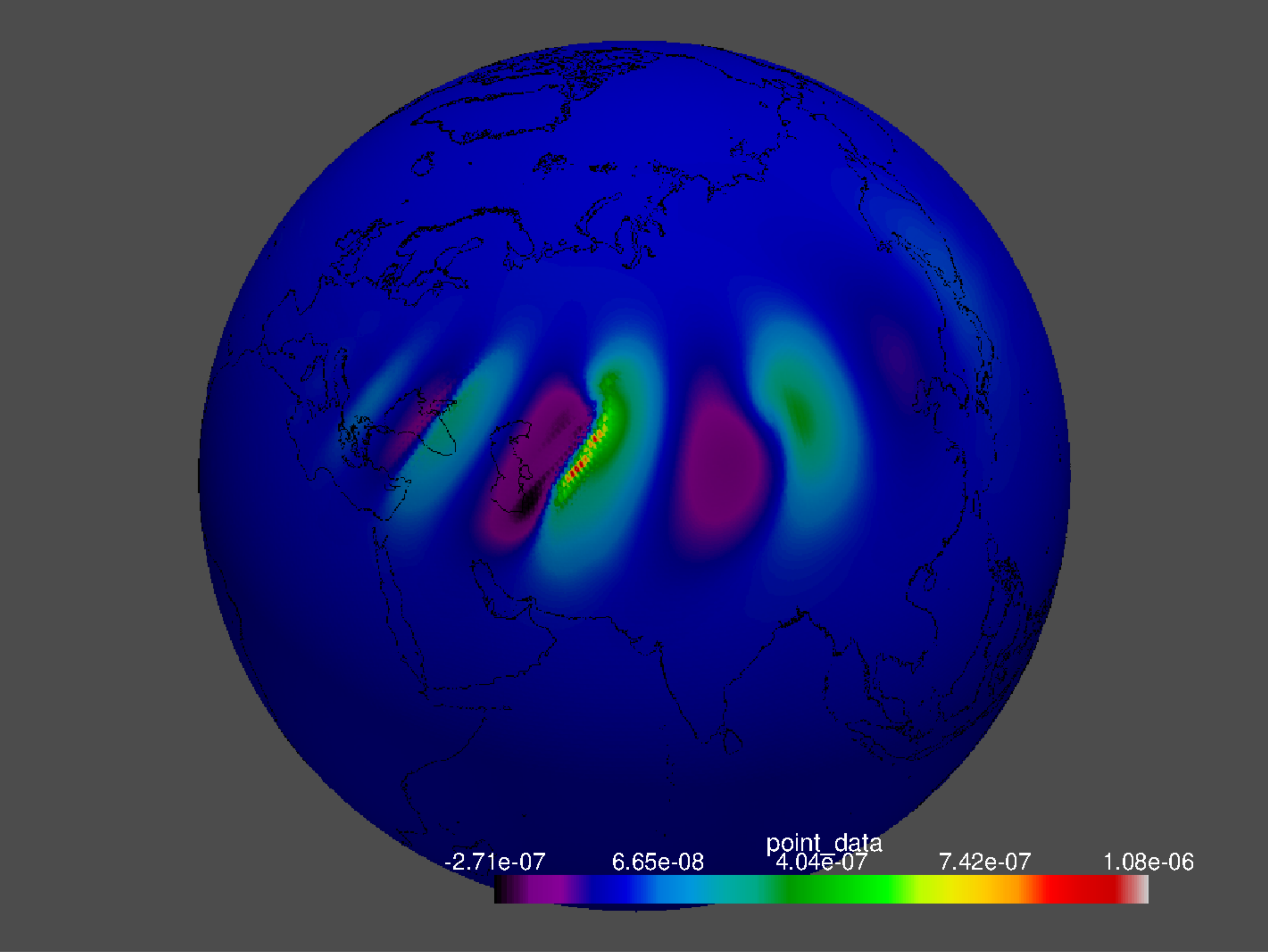}
	\caption{Surface level divergence for the ROS34PRW vector invariant (top) and advective form (bottom) 
	at day 7 for the baroclinic wave test case at the C96 resolution.}
\label{fig::ce_bw_6_2}
\end{center}
\end{figure}

\begin{figure}[!hbtp]
\begin{center}
\includegraphics[width=0.48\textwidth,height=0.36\textwidth]{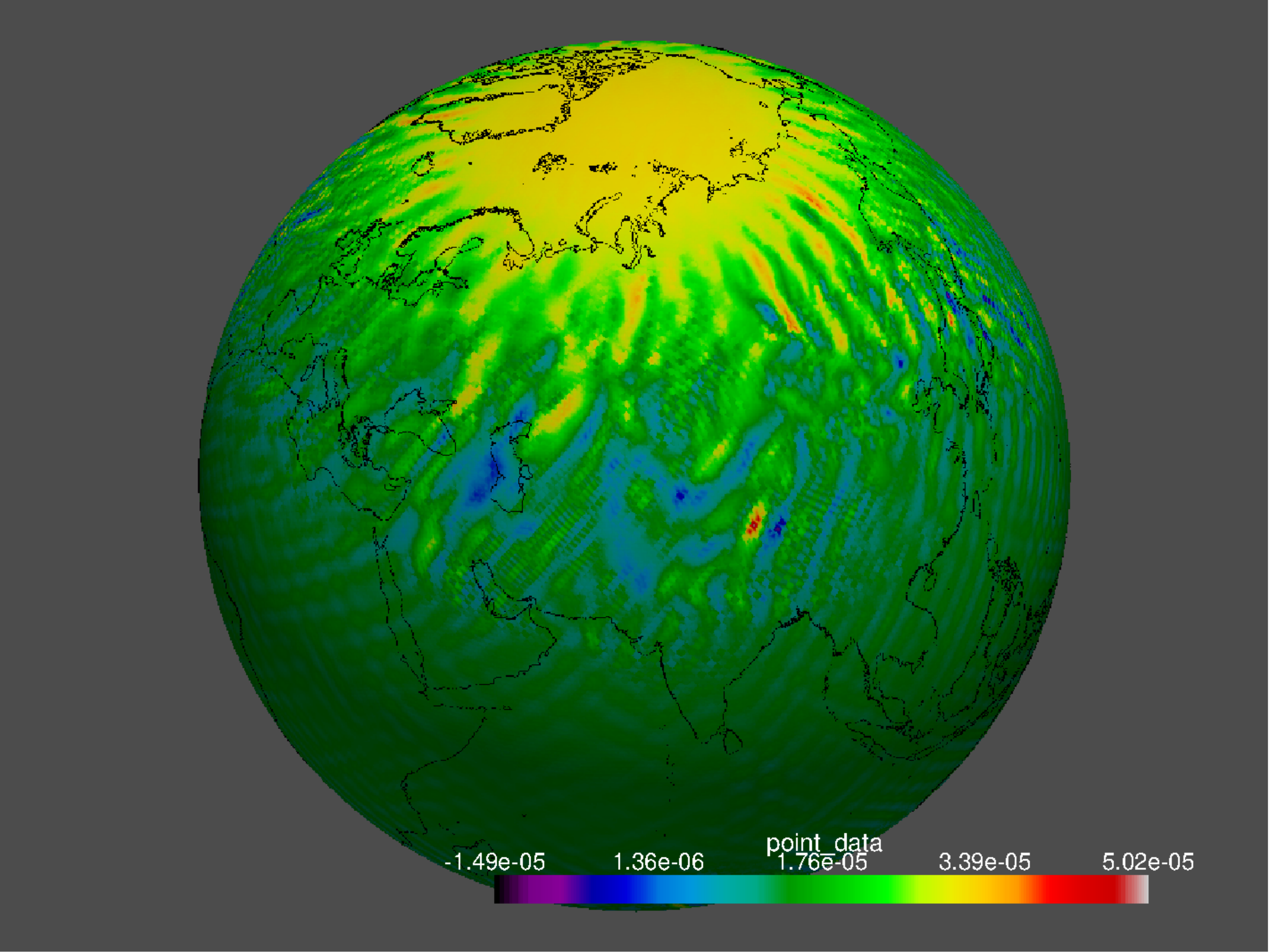}
\includegraphics[width=0.48\textwidth,height=0.36\textwidth]{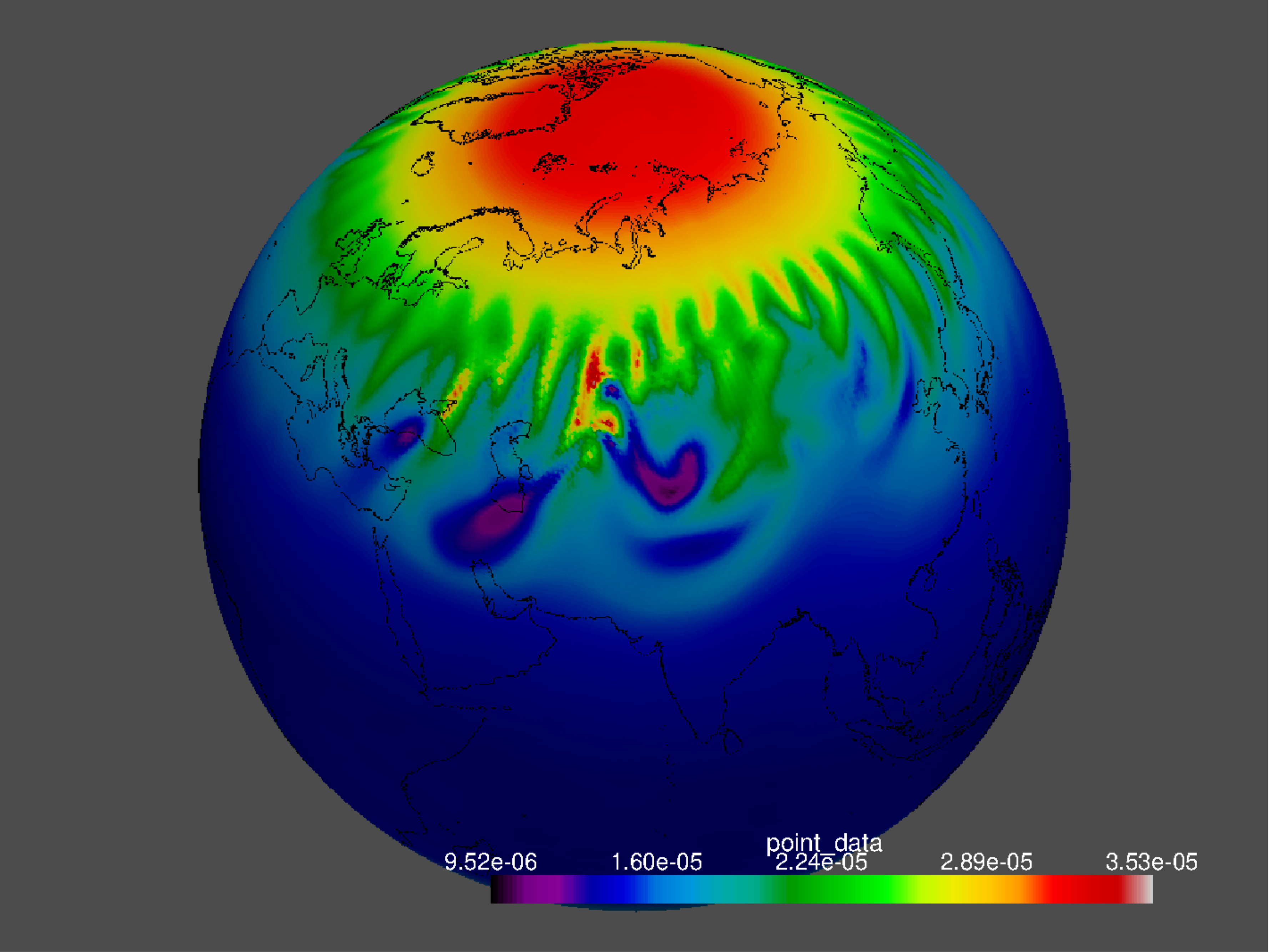}
	\caption{Surface level divergence the CN4 vector invariant (top) and advective form (bottom) 
	at day 7 for the baroclinic wave test case at the C96 resolution.}
\label{fig::ce_bw_6_1}
\end{center}
\end{figure}

Meridional biases are also observed for the CN4 scheme in both advective and vector invariant forms 
for the lowest level \red{three dimensional } divergence as presented in Fig. \ref{fig::ce_bw_6_1}. These biases are an order
of magnitude greater than the divergence associated with the baroclinic instability. \red{The divergence
resulting from the baroclinic process is }
clearly observed without such biases for the ROS34PRW formulations in Fig. \ref{fig::ce_bw_6_2}.

\begin{figure}[!hbtp]
\begin{center}
\includegraphics[width=0.48\textwidth,height=0.36\textwidth]{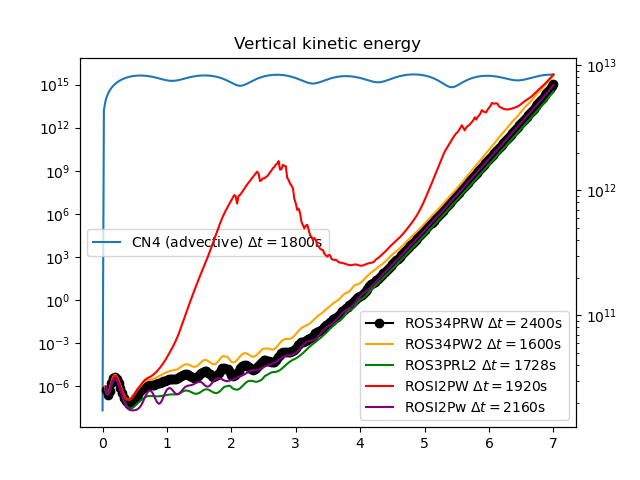}
\includegraphics[width=0.48\textwidth,height=0.36\textwidth]{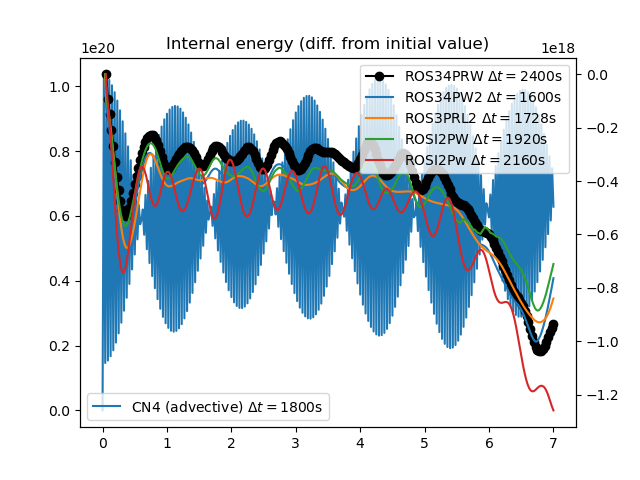}
\includegraphics[width=0.48\textwidth,height=0.36\textwidth]{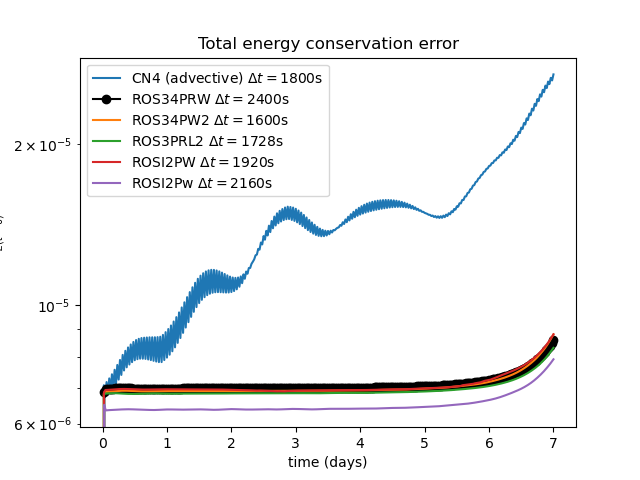}
	\caption{Comparison of the CN4 and RoW schemes at their largest observed 
	stable time step using centered flux reconstructions for the C96 resolution
	with respect to the vertical kinetic energy (top), internal (center), and total 
	energy conservation error (bottom). Note the 
	different scales on the vertical axes for the CN and RoW schemes 
	in the vertical kinetic energy plot.}
\label{fig::ce_bw_7}
\end{center}
\end{figure}

We also compare the energetic profiles for the various RoW schemes studied for 
the shallow water equations in Section 3.1. for the C96 resolution in Fig. \ref{fig::ce_bw_7}. 
These are presented in vector invariant form for the largest time steps for which the different 
schemes were observed to be stable and convergent using centered flux reconstructions.
We exclude results for the ROWDAIND2 and ROS34PW3 schemes, as these were observed to be rapidly 
unstable, which is consistent with the shallow water results, where these schemes produced a 
positive growth in energy as seen in Figs. \ref{fig::sw_1}, \ref{fig::sw_2}.
In the case of the ROS34PW3 scheme, this is perhaps a consequence of the fact that it is 
\red{only A-stable and } not 
\blue{L-stable, meaning that the amplification factor $R$ as discussed in Section \ref{sec::1.1} 
does not go to zero as the eigenvalues approach infinity}.

In all cases the RoW schemes exhibit lower energy conservation error than the CN4 
scheme (in advective form). When comparing the vertical kinetic energy profiles, we see that 
the RoW schemes all exhibit a growth associated with the baroclinic 
instability that is consistent with previous results \cite{Lee (2021)}, and that is two orders of 
magnitude smaller \red{than } that associated with the spurious oscillation of the CN4 scheme. The 
ROS34PRW scheme is observed to be stable and convergent at a time step of $\Delta t=2400s$, 
which is approximately 30\% longer than the maximum observed stable time step for the CN4 scheme
at $\Delta t=1800s$. \green{This result is consistent with those observed for the shallow water 
experiments in Section \ref{sec::galewsky}, where the ROS34PRW was observed to be stable for time 
steps 20\% longer than the CN4 scheme}. In all cases the configurations at the maximum observably 
stable time step required two forward Euler transport sub-steps for the majority of the simulation 
time. 

\begin{figure}[!hbtp]
\begin{center}
\includegraphics[width=0.48\textwidth,height=0.36\textwidth]{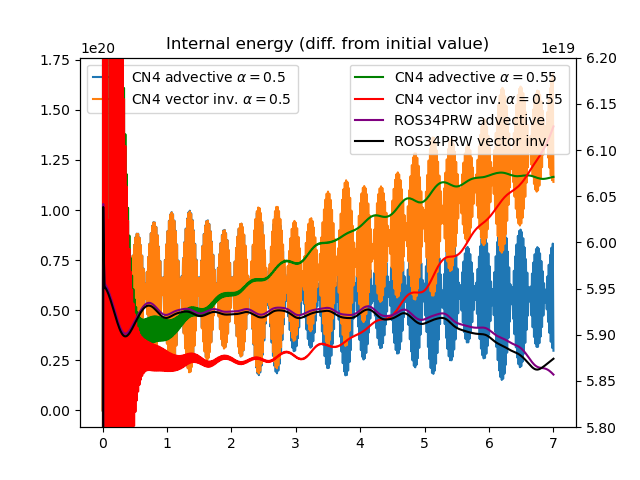}
\includegraphics[width=0.48\textwidth,height=0.36\textwidth]{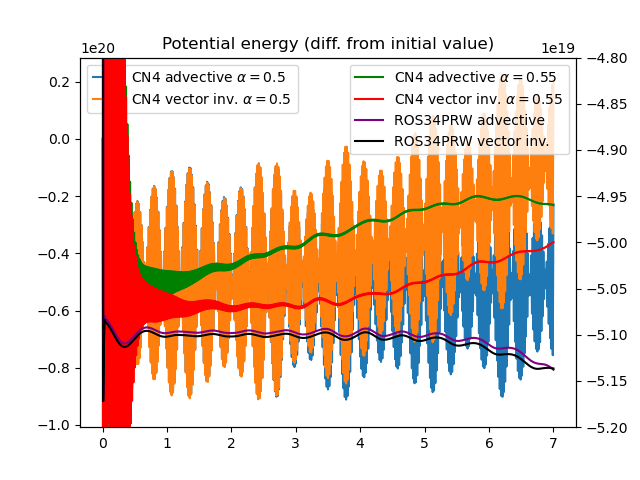}
\includegraphics[width=0.48\textwidth,height=0.36\textwidth]{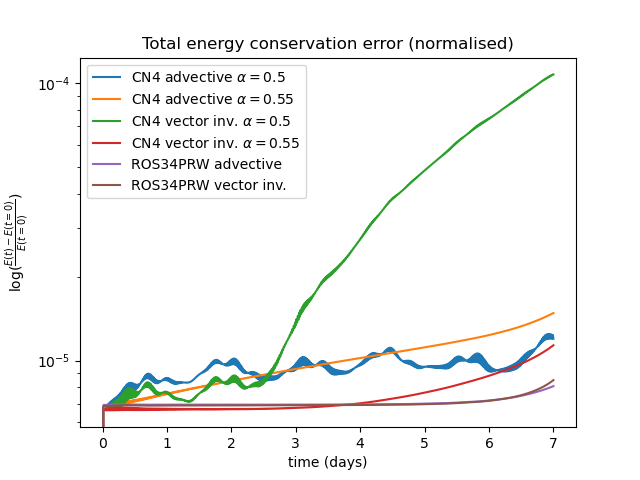}
	\caption{Comparison of the CN4 and ROS34PRW schemes using centered fluxes for the C96 resolution
	with off-centered $\alpha$ with respect to the internal energy (top), potential energy (center) differences from their
	initial values, and total energy conservation error (bottom). Note the difference in scales
	on the vertical axes for the internal and potential energy plots.}
\label{fig::ce_bw_8}
\end{center}
\end{figure}

One method of suppressing the fast oscillation observed for the CN4 scheme is to
off-center the time discretisation in favor of the future time level $n+1$, as given in
\eqref{eq::cn_rhs_alpha}, as this ensures that the time discretisation is no longer neutrally 
stable. The downside of this approach is that it degrades the formal accuract of the method.
In Fig. \ref{fig::ce_bw_8} we compare the internal and
potential energy and energy conservation error profiles for the CN4 scheme using centered 
($\alpha=0.5$) and off-centered ($\alpha=0.55$) time discretisations with respect to the
ROS34PRW scheme for the C96 resolution.

As observed, the off-centering does ultimately suppress the spurious oscillation, such that
it is of smaller amplitude that the long time signal after approximately 1-2 days.
However the energy conservation errors are still greater than for the 
ROS34PRW scheme, which exhibits no such oscillation even at short times. Even with the 
off-centered time discretisation, the CN4 schemes still exhibit spurious growth in the
potential and internal energies, which should actually be decreasing to balance the
growth in kinetic energy of the baroclinic instability as is the case for the ROS34PRW
scheme, and also for previous results using an integrator that exactly conserves energy
for the vertical dynamics \cite{Lee (2021)}.

\begin{figure}[!hbtp]
\begin{center}
\includegraphics[width=0.48\textwidth,height=0.36\textwidth]{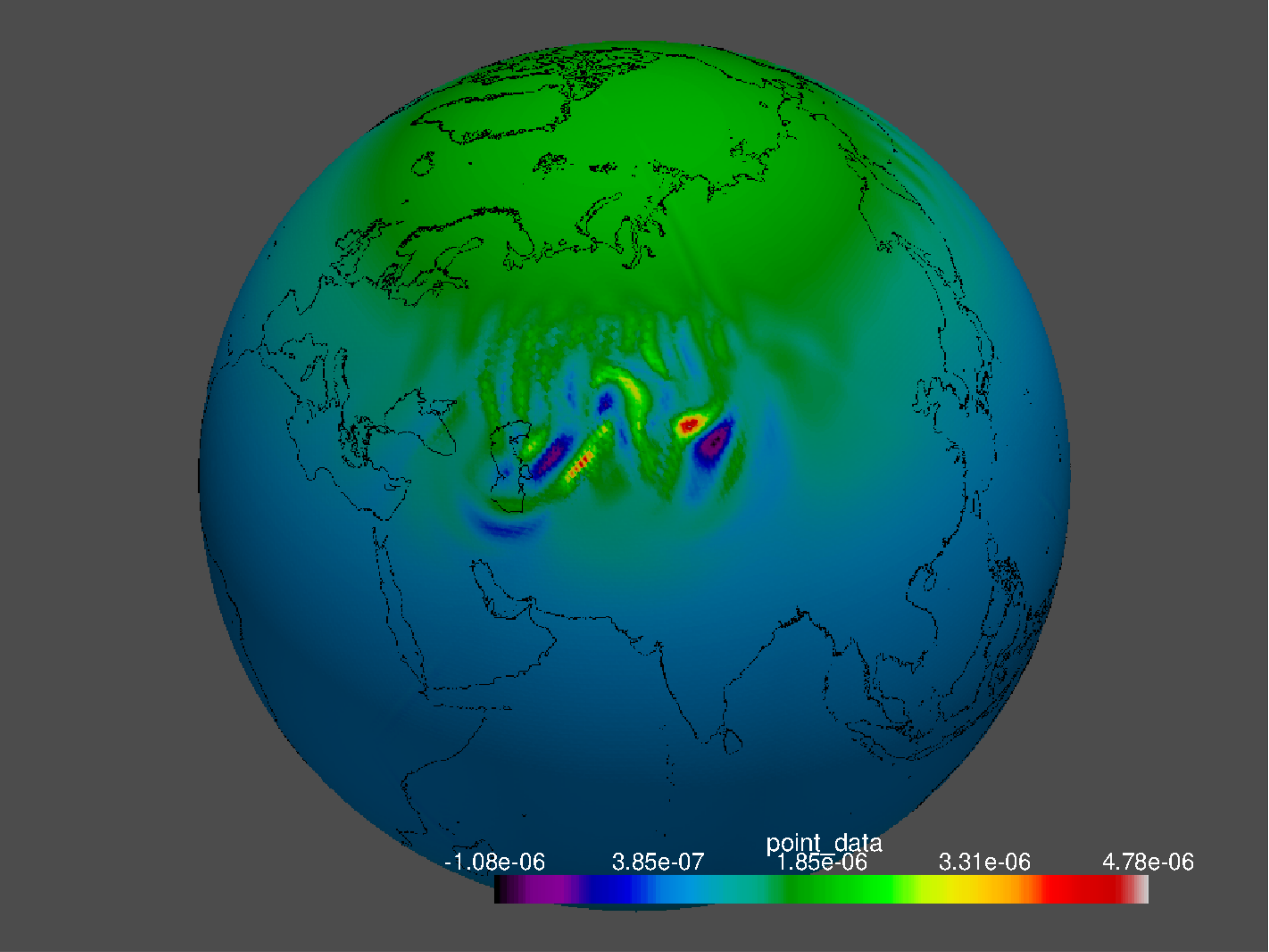}
\includegraphics[width=0.48\textwidth,height=0.36\textwidth]{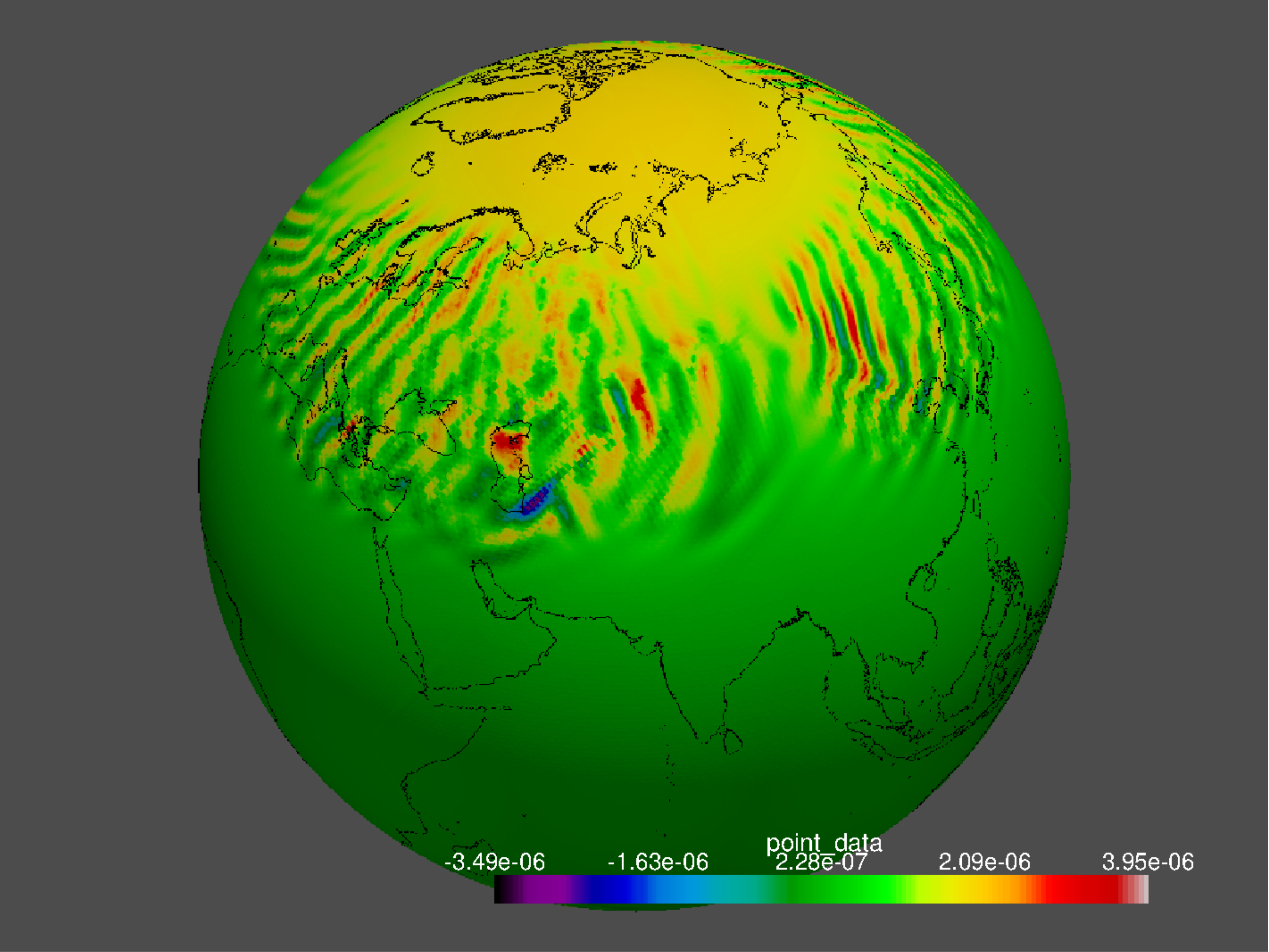}
	\caption{Surface level divergence for the temporally off-centered CN4 advective (top) and 
	vector invariant (bottom) form at day 7 for the baroclinic wave test case at the C96 resolution.}
\label{fig::ce_bw_9}
\end{center}
\end{figure}

We also compare the lowest level divergence for the off-centered advective and vector invariant 
CN4 schemes in Fig. \ref{fig::ce_bw_9}. While the meridional biases are
reduced somewhat compared to those observed for the centered formulations in Fig. \ref{fig::ce_bw_6_1}, 
they are still present and greater than the physical divergence of the baroclinic instability,
which is observed without such bias for the ROS34PRW scheme in Fig. \ref{fig::ce_bw_6_2}.

\subsubsection{\blue{Performance with upwinded transport terms}}\label{sec::bw_time}

In order to compare the computational performance of the different schemes we use an upwinded
reconstruction of the fluxes in the transport terms, which introduces additional internal
dissipation that allows for longer time steps. We focus on the advective form of the momentum
equation such that dissipation is applied to the momentum transport as well. For the CN
scheme we use a value of $\gamma=1$ in the approximate Jacobian for both the density
$\boldsymbol{\mathsf{D}}^{\rho *}$ and potential temperature
$\boldsymbol{\mathsf{P}}^{\theta *}_{\theta u}$ operators in \eqref{eq::ce_W}, in order to 
over-relax the solution of these transport terms, while keeping a value of $\gamma=1/2$ in the 
momentum equation terms.

The transport terms are sub-stepped using a strong-stability preserving $3^{\mathrm{rd}}$ order 
Runge-Kutta (SSP-RK3) scheme at each sub-step, and for the RoW schemes this was applied using
the velocity increments for the transport velocity as in \eqref{eq::row_adv} (with single 
forward Euler steps at the second and third stages and sub-stepped SSP-RK3 on the first and 
fourth stages). 

The CN scheme was tested in three different configurations, firstly using four 
iterations with the transport terms applied on the first and third stages (two-outer/two-inner 
iterations; 2o2i), secondly with the transport applied at each of the four stages, as in Section 
\ref{sec::bw_stab} (four-outer/one-inner iteration; 4o1i), and thirdly using three iterations with 
the transport at every stage (three-outer/one-inner iteration; 3o1i). The four iteration schemes 
were also run using the L-stable CN scheme described in Section \ref{sec::1.1}. We also explored 
the three stage RoW method in \cite{Jahn et. al. (2015)}, however this could not be run stably.

\begin{figure}[!hbtp]
\begin{center}
\includegraphics[width=0.48\textwidth,height=0.36\textwidth]{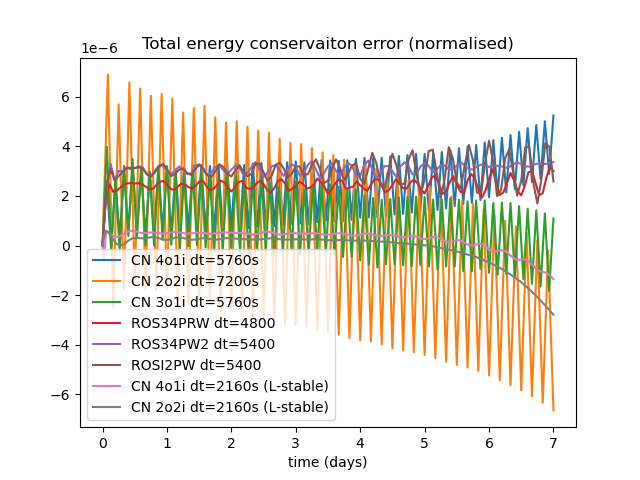}
\includegraphics[width=0.48\textwidth,height=0.36\textwidth]{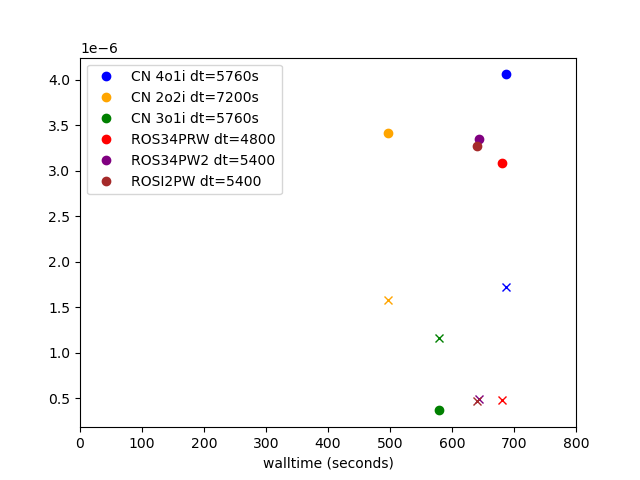}
	\caption{Top: Normalised total energy conservation errors as a function of time for the CN and RoW 
	schemes in advective form with upwinded transport at the maximum stable time step.
	Bottom: Energy conservation error (averaged over the last two time steps; circles) and maximum 
	absolute divergence (crosses) at the last time step as a function of the time to solution for the 
	advective form with upwinded transport at the maximum stable time step.
	L-stable CN variants are excluded.
	}
\label{fig::ce_bw_upwind_1}
\end{center}
\end{figure}

The wall times for 7 days of simulation at the C96 resolution for the upwinded configurations are listed 
in Table II. The CN 2o2i and CN 3o1i schemes allowed for the fastest time to solution, ahead of the RoW 
schemes, with the CN 4o4i behind owing to the application of the sub-stepped SSP-RK3 transport at each
nonlinear iteration. The L-stable ($\gamma=3.1426$) CN schemes were the least efficient, 
owing to the severe time step restriction of these schemes. The $\gamma=0.2716$ L-stable CN schemes were
also explored, however these were not stable even with shorter time steps of $\Delta t=900s$.

\begin{table*}
\begin{center}
\begin{tabular}{|c|c|c|}
	\hline
	Scheme & $\Delta t$ (seconds) & Wall time (seconds) \\
	\hline
	CN 4o1i & 5760.0 & 686.82 \\
	CN 2o2i & 7200.0 & 496.26 \\
	CN 3o1i & 5760.0 & 578.50 \\
	ROS34PRW & 4800.0 & 680.85 \\
	ROS34PW2 & 5400.0 & 644.05 \\
	ROSI2PW & 5400.0 & 640.51 \\
	CN 4o1i ($\gamma=3.1426$) & 2160.0 & 1861.59 \\
	CN 2o2i ($\gamma=3.1426$) & 2160.0 & 1757.60 \\
	\hline
\end{tabular}
\end{center}
	\caption{Wall times for the different schemes at their maximum stable time step
	for 7 days of simulation at the C96 resolution, using upwinded transport in advective form.}
\end{table*}

In order to investigate the energetics of these upwinded configurations the total energy conservation
errors are given in Fig. \ref{fig::ce_bw_upwind_1}. As for the centered schemes, the upwinded CN schemes
exhibit a temporal oscillation with a frequency of $2\Delta t$, which trends towards instability for the
CN 4o1i scheme. To a lesser degree this oscillation is also present for the RoW schemes, suggesting that
their L-stable nature is insufficient to fully damp these internal oscillations in advective form with
large time steps. Only the L-stable CN schemes appear to be free of this oscillation. This figure also 
shows the absolute value of the energy conservation error (averaged over the last two time steps to remove
the $2\Delta t$ oscillation) and the absolute value of the maximum lowest level divergence at day 7 for 
the different schemes as a function of time to solution. While the CN 2o2i scheme is the most efficient, 
it also suffers from larger conservation and divergence errors than the CN 3o1i scheme. The RoW schemes
have the smallest divergence errors but relatively large energy conservation errors, suggesting that
these schemes are somewhat under-damped in advective form. The L-stable CN schemes are not shown, owing
to their significantly larger wall times.

\begin{figure}[!hbtp]
\begin{center}
\includegraphics[width=0.48\textwidth,height=0.36\textwidth]{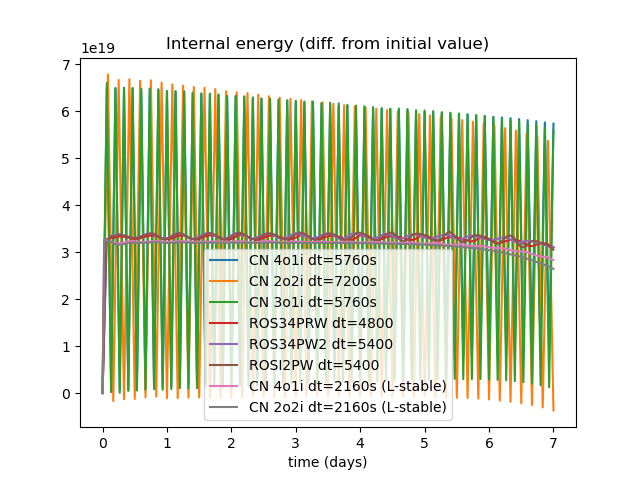}
\includegraphics[width=0.48\textwidth,height=0.36\textwidth]{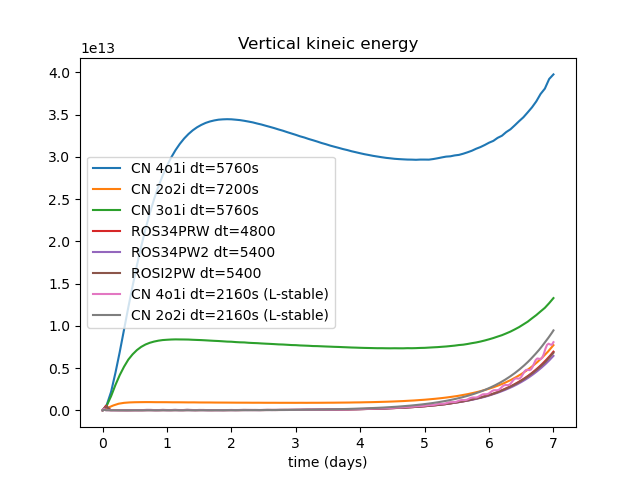}
	\caption{Temporal evolution of the internal energy (difference from initial time) (top)
	and the vertical component of the kinetic energy (bottom) for the CN and RoW schemes in
	advective form with upwinded transport at the maximum stable time step.}
\label{fig::ce_bw_upwind_2}
\end{center}
\end{figure}

The internal energy as a function of time for the upwinded schemes is given in Fig. \ref{fig::ce_bw_upwind_2}.
Unlike for the centered schemes in Fig. \ref{fig::ce_bw_1}, the upwinded CN schemes all exhibit the
correct downward trend in internal (and potential, not shown) energy in order to balance out the
growth in kinetic energy of the baroclinic instability. This figure also shows the vertical kinetic 
energy as a function of time. While the $\mathcal{O}(100)$ error observed for the CN schemes with 
centered fluxes in Fig. \ref{fig::ce_bw_2} is not present for the upwinded schemes, there is still 
a large error observed for the CN 4o1i and CN 3o1i schemes, with the CN 2o2i and L-stable CN schemes 
yielding results consistent with the RoW schemes.

\begin{figure}[!hbtp]
\begin{center}
\includegraphics[width=0.48\textwidth,height=0.36\textwidth]{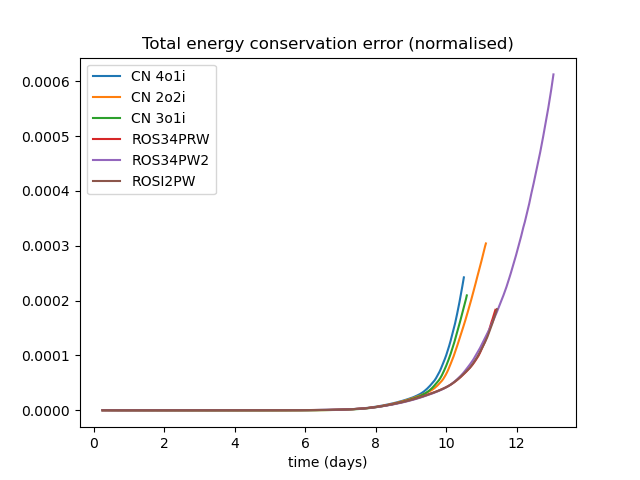}
\includegraphics[width=0.48\textwidth,height=0.36\textwidth]{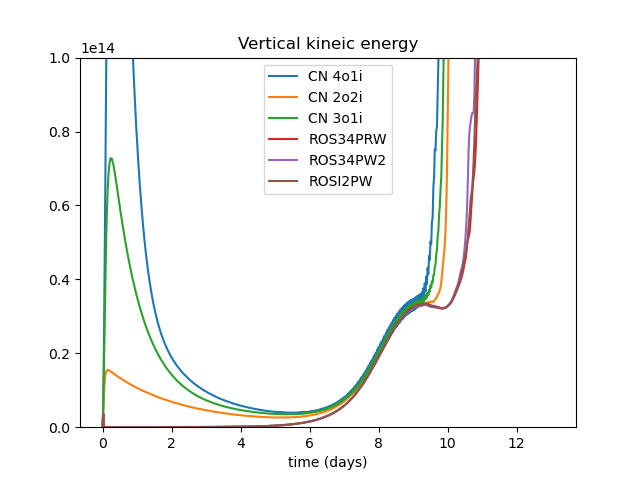}
	\caption{Temporal evolution of the normalised total energy conservation errors (top)
	and the vertical component of the kinetic energy (bottom) for the CN and RoW schemes in
	advective form with upwinded transport for a fixed time step of $\Delta t=1800s$.}
\label{fig::ce_bw_upwind_3}
\end{center}
\end{figure}

Finally, we also show the total energy conservation errors and vertical kinetic energy for the 
upwinded CN and RoW configurations with a fixed time step of $\Delta t=1800s$ in Fig. \ref{fig::ce_bw_upwind_3}.
In all cases the the RoW schemes run for somewhat longer than the CN schemes before also ultimately
going unstable due to insufficient damping. There is also a large perturbation in the initial 
vertical kinetic energy for the CN schemes due to the internal oscillations that is not present
for the RoW schemes.

\subsection{3D compressible Euler: warm bubble on the plane}\label{sec::wb}

While the baroclinic instability test case is a good measure of the 
dynamics \red{at } planetary scales, the non-hydrostatic and compressible dynamics are negligible. 
Consequently we also compare the different schemes for a high resolution
3D warm bubble test case \cite{Giraldo et. al. (2013), Melvin et. al. (2019), Lee (2021), Lee and Palha(2021)} 
for which these effects are significant. The model is configured with an initial state of constant 
density in hydrostatic balance in a three dimensional horizontally periodic box using $100\times 100\times 150$ 
lowest order elements with a uniform resolution of $10m$. 
The balanced state is overlaid with a small potential temperature perturbation, which rises and 
distorts over several minutes. 
\blue{We compare the schemes using the advective form with upwinded flux reconstructions, as in Section 
\ref{sec::bw_time}, with the transport terms applied as in \eqref{eq::row_adv} for the RoW schemes. }
Unlike for the baroclinic test case, monotone transport is applied for the potential temperature.

Table III details the wall times and errors in the maximum value of the potential temperature, $\theta_{max}$
at 400s for the different schemes. In contrast to the baroclinic test case, here the RoW schemes
allow for longer time steps and shorter times to solution. Despite the application of longer time 
steps, the errors in $\theta_{max}$ are also smaller for the RoW schemes. The ability to run the 
RoW schemes for the longer time steps than the CN schemes at the non-hydrostatic scale is perhaps
due to the L-stable nature of the RoW schemes, and their corresponding capacity damp fast oscillations associated
with initial hydrostatic imbalance and acoustic modes.

\begin{table*}
\begin{center}
\begin{tabular}{|c|c|c|c|}
	\hline
	Scheme & $\Delta t$ (seconds) & Wall time (seconds) & $L_{\infty}$ error in $\theta_{max}$ \\
	\hline
	CN 4o1i  & 0.8  & 3780.32 & 0.053841 \\
	CN 2o2i  & 0.8  & 3318.10 & 0.049282 \\
	CN 3o1i  & 0.8  & 3810.97 & 0.053183 \\
	ROS34PRW & 1.33 & 2661.80 & 0.026015 \\
	ROS34PW2 & 1.33 & 2841.93 & 0.026591 \\
	ROSI2PW  & 1.33 & 2955.06 & 0.026119 \\
	\hline
\end{tabular}
\end{center}
	\caption{Wall times and error in the maximum value of the potential temperature, $\theta_{max}$, 
	at 400s for the different CN and RoW schemes at their maximum stable time step
	for the warm bubble test case using upwinded transport in advective form.}
\end{table*}

\begin{figure}[!hbtp]
\begin{center}
\includegraphics[width=0.48\textwidth,height=0.36\textwidth]{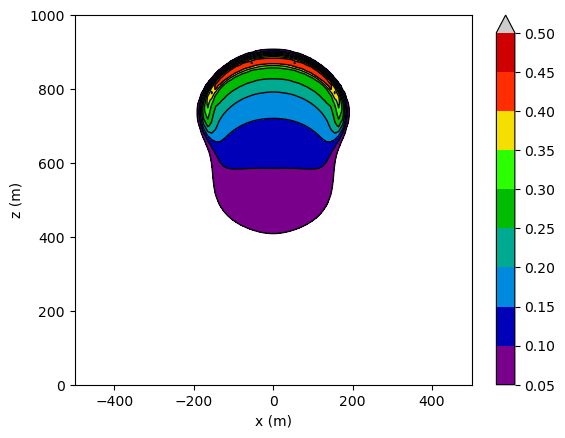}
\includegraphics[width=0.48\textwidth,height=0.36\textwidth]{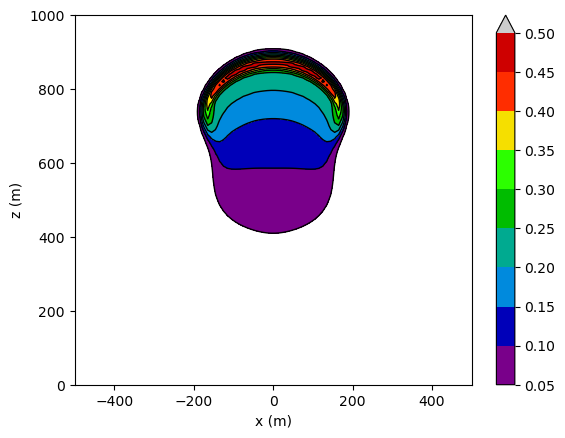}
	\caption{Potential temperature cross section at $y=0$ for the 3D warm bubble test case at time $400s$ for 
	\blue{the CN 2o2i scheme $\Delta t=0.8s$ (top) and the ROS34PRW 
	scheme, $\Delta t=1.33s$ (bottom), both in advective form with upwinded flux reconstruction.}}
\label{fig::ce_wb_1_2}
\end{center}
\end{figure}

\begin{figure}[!hbtp]
\begin{center}
\includegraphics[width=0.48\textwidth,height=0.36\textwidth]{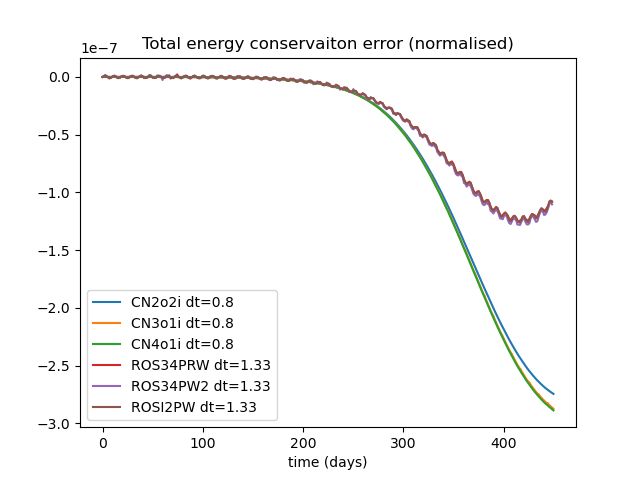}
\includegraphics[width=0.48\textwidth,height=0.36\textwidth]{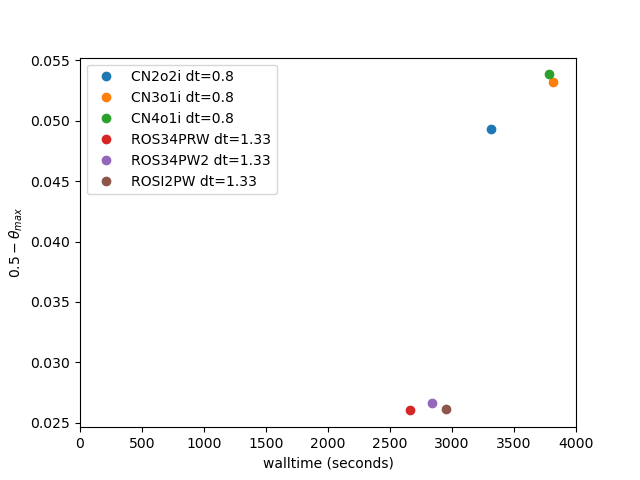}
	\caption{Normalised total energy conservation errors (top) and error in the maximum 
	value of potential temperature as a function of wall time (bottom) at maximum stable
	time step 
	with upwinded flux reconstruction for the warm bubble test case.}
\label{fig::ce_wb_2}
\end{center}
\end{figure}

The less diffusive nature of the RoW schemes is also observed in the potential temperature cross sections
at 400s in Fig. \ref{fig::ce_wb_1_2}, which compares the most accurate RoW (ROS34PRW) and CN (2o2i) schemes.
The normalised energy conservation errors are presented in Fig. \ref{fig::ce_wb_2}. While the CN schemes
display a continued decay in total energy, this begins to increase as the bubble ascends for the RoW schemes.
This tracks closely with the internal energy evolution of the different schemes (not shown), which also show
a steady decrease for the CN schemes and an uptick for the RoW schemes. 
Previous results \cite{Lee (2021)} using an exact energy conserving vertical integrator suggest
that while the potential energy should exhibit a negative trend as the bubble ascends, the 
mean internal energy (aside from the fast oscillation) should stay relatively constant, so
from this we infer that the RoW schemes are more \green{physically realistic}.

This figure also shows a scatter plot detailing the errors in $\theta_{max}$ at 400s as a function of time
to solution at maximum stable time step, as also presented in Table III. 
The RoW schemes are both the most efficient as well as the most accurate by this measure. This is in 
contrast to the similar plot in Fig. \ref{fig::ce_bw_upwind_1}, where the CN2o2i and CN3o1i schemes are
the most efficient for the baroclinic wave test case.

\section{Conclusions}

This article compares the results for a collection of four stage RoW time 
integrators with an approximate Jacobian as a substitute for
a CN time discretisation for the solution of geophysical systems 
at planetary and non-hydrostatic scales. Comparisons are made in terms of both energetics 
for a vector invariant formulation with centered fluxes, in order to minimise the amount 
of internal dissipation, \blue{and also in terms of time to solution for an advective formulation 
with upwinded fluxes, in order to introduce internal dissipation and allow for longer time
steps and shorter wall times.}

For the vector invariant form with centered fluxes, the best performing RoW schemes allow
for longer time steps than the four iteration CN scheme for both the shallow water and 3D 
compressible Euler equations at planetary scales. Also for the 3D compressible Euler equations
the RoW schemes are free of the spurious high frequency oscillations, divergence errors and 
upward trends in internal and potential energy that are present for the CN scheme, even with
off-centering in the temporal discretisation.

For the upwinded advective formulations the results are more mixed, with the CN 2o2i scheme 
(four iterations with transport terms applied on the first and third iterations) allowing 
for the longest time steps and shortest wall times for the planetary scale baroclinic instability
test case, and the RoW schemes allowing for the longest time steps and shortest wall times 
for the non-hydrostatic rising bubble test case. The other CN schemes (3o1i and 4o1i) exhibited
large errors in the vertical kinetic energy despite also being more performant than the RoW 
schemes at planetary scales, and at non-hydrostatic scales the RoW methods were also less
diffusive than the CN schemes as well as being more performant.

We also introduce a new temporal formulation of the transport terms for the RoW schemes, based
on the transport of solution increments at subsequent stages, that allowed for the efficient
implementation of the RoW methods in advective form. An L-stable variant of the four stage 
CN scheme was also derived. While this effectively cleaned up the spurious oscillations of the
CN scheme in advective form, it was shown to be stable only for small time steps that limited
its performance, and was not observed to be stable with centered fluxes.

\red{Finally, we note that the results of this study will be somewhat sensitive to the choice
of approximations made in the evaluation of the Jacobian. While modifications to the Jacobian
approximation for the shallow water model yielded little variation in results, there may be
more variation for LFRic, for which the presence of acoustic modes and complex thermodynamics
may introduce additional sensitivity.}

\section{Acknowledgments}

David Lee would like to thank Drs. Gary Dietachmayer and Nigel Wood for their helpful 
comments and insights on an early version of this manuscript, and Drs. Thomas Melvin 
and Ben Shipway for their constructive comments and advice. David is also grateful
to the three anonymous reviewers, all of whom provided helpful feedback that improved
this manuscript.

The author acknowledges no conflict of interest.

\section{CRediT authorship contribution statement}
David Lee: Conceptualisation, data curation, formal analysis, investigation, methodology, 
project administration, software, validation, visualisation, writing: reviewing and editing.

\end{document}